\documentclass[gmd, manuscript]{copernicus}

\usepackage{silence}
\usepackage{amsthm}
\usepackage{float}
\usepackage{subcaption}
\usepackage[]{graphicx}
\usepackage{pgfplots}
\pgfplotsset{compat=newest}
\usepackage{adjustbox}
\usepackage{xurl}
\usepackage{xspace}
\usepackage[utf8]{inputenc}
\usepackage{tikz}
\usetikzlibrary{trees}
\usepackage[edges]{forest}
\usetikzlibrary{arrows.meta}
\nolinenumbers
\usepackage{booktabs}
\usepackage{amsmath}
\newcommand{\plotbox}[1]{#1}
\newcommand{\ignore}[1]{}

\newcommand{\Sec}[1]{Section \ref{sec:#1}}

\newcommand{\Fig}[1]{Figure \ref{fig:#1}}

\begin{document}

\title{The Ice Sheet State and Parameter Estimator (ICESEE) Library (v1.0.0): Ensemble Kalman Filtering for Ice Sheet Models}



\Author[1]{Brian}{Kyanjo}
\Author[2]{Talea L.}{Mayo}
\Author[1]{Alexander A.}{Robel}

\affil[1]{School of Earth and Atmospheric Sciences, Georgia Institute of Technology, Atlanta, GA, USA}
\affil[2]{Department of Mathematics, Emory University, Atlanta, GA, USA}




\correspondence{Brian Kyanjo (bkyanjo3@gatech.edu)}

\runningtitle{The Ice Sheet State and Parameter Estimator (ICESEE) Library (v1.0.0): Ensemble Kalman Filtering for Ice Sheet Models}

\runningauthor{Brian Kyanjo, Talea L. Mayo, Alexander A. Robel}

\received{}
\pubdiscuss{} 
\revised{}
\accepted{}
\published{}


\firstpage{1}

\maketitle
\begin{abstract}
ICESEE (ICE Sheet statE and parameter Estimator) is a Python-based, open-source data assimilation framework designed for seamless integration with ice-sheet and other Earth system models. ICESEE implements a parallel Ensemble Kalman Filter (EnKF) architecture with full Message Passing Interface (MPI) support, enabling scalable assimilation in both model state and parameter spaces. Its core algorithm employs a nonlinear, transformation-based update scheme originally proposed by \cite{evensen:2003}, which avoids explicit construction of the forecast error covariance matrix. This matrix-free formulation eliminates the need for localization while retaining robustness in high-dimensional, nonlinear systems. In addition to the core EnKF, ICESEE provides serial implementations of four alternative EnKF variants, including a localized formulation, offering flexibility for methodological testing and comparative studies. Beyond state estimation, ICESEE supports the indirect inference of unobserved or weakly constrained model parameters through a hybrid assimilation–inversion strategy. In this approach, ensemble-based data assimilation corrects the model state using available observations, while physics-based inverse methods are subsequently applied to infer parameters such as basal friction. The framework features modular coupling interfaces, adaptive state indexing, and efficient parallel input/output, making it readily extensible to a wide range of numerical modeling environments. ICESEE has been successfully coupled with several existing models, including the MATLAB/C++-based Ice Sheet and Sea-level System Model (ISSM), the Firedrake-based Python model Icepack, a one-dimensional flowline model, and the reduced-order Lorenz--96 system. In this study, we focus on applications with ISSM and Icepack to demonstrate ICESEE’s interoperability, numerical performance, scalability, and its ability to jointly improve state estimates and infer uncertain model parameters. Performance benchmarks show strong and weak scaling on high-performance computing platforms, underscoring ICESEE’s potential to enable long-term, observation-constrained reanalyses of ice-sheet evolution at continental scales.
\end{abstract}


\section{Introduction}
\label{sec:intro}
Ice sheets play a central role in Earth's climate system, regulating global sea levels and influencing broader atmospheric and oceanic circulation patterns \citep{nowicki2016, shepherd2018mass}. Accurate simulation of ice sheet dynamics is therefore essential for projecting sea level rise and assessing the long-term impacts of climate change. However, traditional ice sheet models often rely on static or poorly constrained parameterizations that are typically not transiently calibrated to sparse observational datasets \citep{goelzer2020design, aschwanden2019}. These limitations introduce considerable uncertainties in projections, reducing the reliability and utility of model-based forecasts.

To address these challenges, data assimilation (DA) methods, which systematically modify model states and parameters to be closer to observations, have been utilized to enhance model accuracy and predictive capability \citep{Evensen2009, carrassi2018data}. Among the available DA techniques, the Ensemble Kalman Filter (EnKF) has gained widespread adoption due to its capacity to be integrated with high-dimensional and nonlinear models without the need to modify the model itself \citep{Evensen1994, gillet2020assimilation, youngmin:2025}. The EnKF offers a principled statistical framework for fusing model forecasts with observations, enabling joint estimation of state and parameter variables and thus improving the fidelity and robustness of model outputs.

Several studies have demonstrated the potential of EnKF to advance ice sheet modeling. \cite{gillet2020assimilation} demonstrated that EnKF can be used to assimilate surface elevation and velocity data in a transient flowline (i.e., intermediate complexity) model of an idealized marine-terminating glacier, effectively constraining basal conditions and enhancing grounding line migration forecasts. \cite{bonan:hal-00837845} applied EnKF ensemble methods to initialize a flowline shallow ice sheet model by incorporating bedrock and surface topography, surface velocities, and elevation trends. Their results highlighted that aligning model initial conditions with current observations is essential for producing reliable future predictions. \cite{bonan:2014} further introduced the Ensemble Transform Kalman Filter (ETKF), a variant of EnKF, to jointly estimate bedrock topography, ice thickness, and basal sliding parameters for a shallow ice flowline model. Similarly, \cite{hossain:2023} applied ETKF to assimilate ice surface elevation and lateral ice extent observations by updating the level set function that describes the ice interface in a flowline model of a marine-terminating glacier. Their work demonstrated the effectiveness of data assimilation for monitoring seasonal and multi-year glacier advance and retreat cycles. By incorporating remotely sensed surface elevation profiles, they achieved more accurate tracking of migrating glacier termini and changes in glacier surface characteristics. More recently, \cite{youngmin:2025} coupled an EnKF-based method with the Ice Sheet System Model (ISSM) \citep{issm} to develop a framework aimed at improving state and parameter estimation for a semi-idealized domain in a two-dimensional ice sheet model widely used for projections (i.e., high complexity). Their study demonstrated that ensemble-based data assimilation methods can robustly recover both basal conditions and model states after only a few assimilation cycles. Furthermore, they observed that assimilating a larger set of observations systematically enhances the accuracy of these estimates, leading to more reliable model projections.

Despite the strengths of the EnKF-based approach, the practical application of the EnKF in ice sheet modeling faces several persistent challenges. One of the most significant issues is ensemble undersampling, where limited ensemble sizes are insufficient relative to the high dimensionality of model state vectors. Additionally, running computationally intensive model simulations to generate ensemble realizations during the forecast step leads to substantial computational costs. Parallel scalability remains problematic, particularly when coupling data assimilation packages with slowly evolving processes such as thermodynamics and bedrock adjustment. Furthermore, observational data are often limited and sparse, exacerbating estimation uncertainties \citep{gillet2020assimilation, goelzer:2017, choi:2023}. These factors collectively constrain the broader applicability and impact of EnKF-based methods. A further major bottleneck identified in the literature is the absence of a standardized, extensible EnKF software library capable of integrating with a diverse range of ice sheet models and assimilation workflows. Existing implementations are frequently tightly coupled to specific model architectures or lack modularity, posing challenges to interoperability, reproducibility, and collaborative development.

To address these limitations, we present the ICE Sheet StatE and parameter Estimator (ICESEE), an open-source, parallel, Python-based, and extensible data assimilation library specifically designed for ice sheet models. ICESEE provides an adaptive, modular interface equipped with standardized application programming interfaces (APIs), enabling seamless coupling with a range of modeling frameworks. Its architecture is model-agnostic and incorporates robust Message Passing Interface (MPI)-based parallelism alongside adaptive mechanisms for both state and parameter estimation. At its core, ICESEE implements a matrix-free update strategy based on the nonlinear formulation introduced by \cite{evensen:2003}, which eliminates the need for explicit covariance matrix computation and facilitates efficient parallel execution. In addition to this primary algorithm, ICESEE includes four serial EnKF variants for benchmarking and experimentation, offering flexibility for a variety of use cases. ICESEE addresses key challenges identified in the literature through several innovations: it supports bounded stochastic updates for reliable parameter estimation, incorporates adaptive inflation and localization to mitigate filter divergence, and features scalable parallelization strategies suitable for computationally expensive, high-resolution models. Moreover, it enables dynamic assimilation of observational data and flexible configuration of ensemble priors, effectively balancing computational and modeling requirements.

We validate the ICESEE framework with a range of ice sheet models, including ISSM \citep{issm}, the Firedrake-based Icepack model \citep{icepack}, a 1D flowline model \citep{Robel2021s}, and a reduced-order Lorenz-96 system \citep{lorenz96}. For this paper, we focus specifically on the ISSM and Icepack models as primary case studies. Results from these two applications demonstrate ICESEE's capability to deliver accurate state and parameter estimations across varying modeling scales and computational platforms.

A defining feature of ICESEE is its focus on standardization and interoperability. By offering a common platform for data assimilation, ICESEE supports reproducible research, enables cross-method comparisons, and encourages collaborative development. Its model-agnostic design and open-source distribution on GitHub \citep{icesee_git} position ICESEE as a valuable resource for both theoretical advancement and practical application within the Earth system modeling community.

\section{The Ensemble Kalman Filter}
\label{sec:EnKF}
Ensemble Kalman Filters (EnKFs) are a class of related data assimilation algorithms designed to efficiently estimate the evolving state of complex systems. They operate by simulating an ensemble of model realizations and updating these realizations using available observational data. ICESEE incorporates multiple EnKF variants to address different computational limitations and modeling needs. Here we will briefly describe certain mathematical and comptuational elements of the EnKF, but for a detailed discussion of the EnKF and its relationship to other data assimilation methods, we direct the interested reader to the original EnKF studies \citep{Evensen1994, evensen:2003} and subsequent reviews \citep{carrassi2018data, ahmed2020}.

\begin{figure}[htbp]
  \centering
  \plotbox{\includegraphics[width=0.75\textwidth,clip=true,trim=0cm 3.0cm 0cm 3cm]{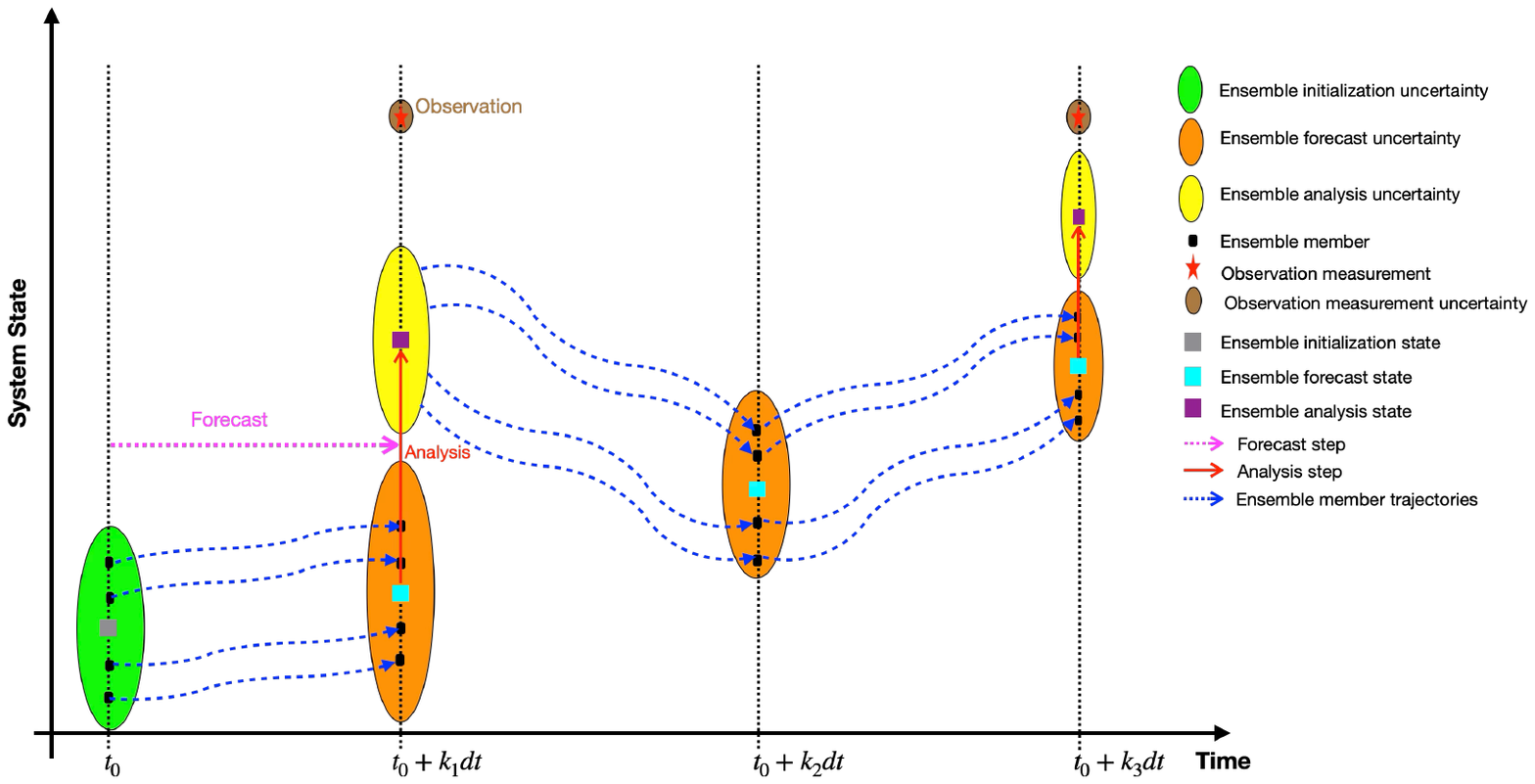}}
  \caption{Schematic of the Ensemble Kalman Filter (adapted from \cite{gillet2020assimilation}) illustrating four assimilation cycles. The process begins with four ensemble members (black dots within the green oval) initialized at time $t_0$, after which the system state (green oval) is advanced through a forecast phase (indicated by the magenta dashed arrow), producing a predicted ensemble state (cyan box within the orange oval). Upon receiving an observation (red star), the state is updated to a new state (purple box inside the yellow oval) via the analysis step (red arrow). The cycle repeats for subsequent forecast and analysis phases, with blue dotted arrows indicating the trajectories of the individual ensemble realizations. Here, $k$ and $dt$ denote the timestep index and time increment, respectively.}
  \label{fig:da_snapshot}
\end{figure}

EnKFs can generally be categorized into two main types: stochastic and deterministic. Both variants are implemented in ICESEE and follow a two-step assimilation cycle: forecasting and analysis (see \Fig{da_snapshot}). The forecast step is consistent across both types. During this step, the model equations are used to advance the model state across $N_e$ ensemble members. This process yields an ensemble mean forecast $\bar{X}^f$, defined as:
\begin{equation}
\bar{X}^f = \frac{1}{N_e} \sum_{i=1}^{N_e} X_i^f ,
\end{equation}
where $X_i^f \in \mathbb{R}^{n \times N_e}$ is the model state vector of size $n$ for each ensemble member $i$. The corresponding forecast error covariance matrix $P^f \in \mathbb{R}^{n \times n}$ is calculated as
\begin{equation}
P^f = \frac{1}{N_e - 1} \sum_{i=1}^{N_e} (X_i^f - \bar{X}^f) (X_i^f - \bar{X}^f)^T .
\label{eqn:cov_model}
\end{equation}
Data to be assimilated are denoted by $y \in \mathbb{R}^m$, a vector of observations with dimension $m$,xxs available at a given analysis time $t^a$. These observations may include, for example, surface velocity, ice thickness, or surface elevation, depending on the specific application. 

\subsection{Stochastic Ensemble Kalman Filter (EnKF)}
\label{sec:EnKF}
Stochastic ensemble filters, such as the EnKF, are generally more effective than deterministic filters (\Sec{deterministic}) at handling nonlinearity in large ensembles. This advantage arises because the inclusion of Gaussian perturbations in the observation vectors mitigates the growth of higher-order non-Gaussian moments in the ensemble generated by nonlinear system dynamics, which degrade the accuracy of ensemble state estimates \citep{Evensen2009}.

In the stochastic EnKF formulation implemented in ICESEE, each ensemble member’s observation vector is independently perturbed to counteract the tendency of the analysis step to underestimate ensemble spread. ICESEE employs a serial implementation inspired by \cite{Evensen1994}, \cite{burger:1998}, and \cite{Evensen2009}. At $t^a$, the updated analysis ensemble is given by: 
\begin{equation}
X_i^a = X_i^f + K (d_i - H(X_i^f)) \quad \forall~ i \in [1, N_e] ,
\label{eqn:SEnKF}
\end{equation}
where $H: \mathbb{R}^n \rightarrow \mathbb{R}^m$ is the observation operator, and $d_i = y^a + \Upsilon _i$ are the ensemble virtual observations at time $t^a$ with $\Upsilon _i$ representing pseudo-random observation errors sampled from a multivariate normal distribution with zero mean and observation error covariance $R$. $K$ is the Kalman gain matrix defined as:
\begin{equation}
K = P^f H^T (H P^f H^T + R_e)^{-1} ,
\label{eqn:kalmann_EnKF}
\end{equation}
where $R_e = DD^T$ is the error covariance matrix with $D \in \mathbb{R}^{m \times N_e}$ being the matrix of ensemble observation perturbations, with columns $d_i ~ \forall~ i \in [1, N_e]$.

For large systems, explicitly computing $P^f$ and inverting the matrix in Equation \eqref{eqn:kalmann_EnKF} becomes computationally expensive. ICESEE adopts a more practical formulation following \cite{evensen:2003}, which avoids direct computation of $P^f$. In this formulation, $P^f$ is expressed as:
\begin{equation}
P^f = \frac{1}{N_e - 1} X' (X')^T ,
\end{equation}
where $X' = X^f - \bar{X}^f$. Similarly, $R_e$ is given by:
\begin{equation}
R_e = \frac{1}{N_e - 1} \Upsilon  \Upsilon ^T .
\end{equation}
Rewriting Equation \eqref{eqn:kalmann_EnKF}, the Kalman gain becomes:
\begin{equation}
K = X' (X')^T H^T (H X' (X')^T H^T + \Upsilon  \Upsilon ^T)^{-1} .
\end{equation}
Assuming uncorrelated observation perturbations, the term $H X' (X')^T H^T + \Upsilon  \Upsilon ^T$ simplifies to $(H X' + \Upsilon ) (H X' + \Upsilon )^T$. This reduces the size of the matrix inversion from  $m^2$ to $mN_e$, making it much more computationally manageable for typical cases where $m>>N_e$.

The singular value decomposition (SVD) of  $(H X' + \Upsilon ) (H X' + \Upsilon )^T = U \Sigma \Sigma^T U^T$, where $U$ is an orthogonal matrix and the diagonal matrix $\Sigma \Sigma^T$ corresponds to the upper $N_e \times N_e$ block of the diagonal matrix $\Lambda$ from:
\begin{equation}
\text{SVD}(H X' (X')^T H^T + \Upsilon  \Upsilon ^T) = Z \Lambda Z^T .
\end{equation}
The Kalman gain can now be written more simply as:
\begin{equation}
K^* = X' (H X')^T (U \Lambda^{-1} U^T)
\label{eqn:kalmann_EnKF_advanced}
\end{equation}
with $\Lambda^{-1}$ having non-zero elements only on its diagonal. Substituting Equation \eqref{eqn:kalmann_EnKF_advanced} into the analysis step from Equation \eqref{eqn:SEnKF} yields the following update expression:
\begin{equation}
X_i^a = X_i^f \bold{\mathcal{X}^{*}}
\label{eqn:_EnKF_advanced}
\end{equation}
Here, $\bold{\mathcal{X}^{*}} = I + K^*_i (d_i - H(X_i^f))$ represents a $N_e \times N_e$ matrix that is both computationally inexpensive to calculate and memory-efficient to store. This matrix formulation enables a scalable and parallel-friendly update mechanism, which is fully integrated into the ICESEE framework. For detailed derivations and further explanation of this approach, see \cite{evensen:2003}.  We choose to use this practical filter formulation only for the results described later in this paper.

\subsection{Deterministic Ensemble Kalman Filters}
\label{sec:deterministic}
There are several deterministic EnKF variants, each distinguished by the structure of the analysis step, particularly in how the Kalman gain is applied. Unlike stochastic methods, deterministic EnKFs avoid perturbing observations for each ensemble member. This eliminates the additional sampling noise introduced by observation perturbations, leading to a more stable estimate of the posterior covariance and reduced computational cost, especially for large ensembles \citep{xuguang:2003, Whitaker:2002}. ICESEE implements the serial versions of several variants described in this section.

\subsubsection{Ensemble Square Root Kalman Filter (EnSRF)}
\label{sec:EnSRF}
The EnSRF variant implemented in ICESEE eliminates the need for perturbing observations, offering advantages in both accuracy and computational efficiency under certain conditions. Instead of relying on stochastic updates, EnSRF applies deterministic transformations involving the square root of the forecast error covariance matrix. This approach determines how the ensemble should be transformed to maintain consistency with the analysis error covariance structure of the Kalman filter.

In EnSRF, the ensemble mean \eqref{eqn:ensrf_mean} is updated using the traditional Kalman gain, while deviations from the mean are adjusted using a reduced Kalman gain. This avoids applying the full Kalman gain to each ensemble member individually, reducing computational cost. By performing these updates deterministically, EnSRF minimizes the randomness present in other ensemble methods, resulting in a more controlled and consistent assimilation process. For a detailed theoretical background, see \cite{tippett:2003} and \cite{Whitaker:2002}. The analysis ensemble mean is computed using:
\begin{equation}
\bar{X}^a = \bar{X}^f + K_e (y^a - H(\bar{X}^f)),
\label{eqn:ensrf_mean}
\end{equation}
where the reduced Kalman gain $K_e = X'(HX')^TS^{-1}$, and $S = HX'(HX')^T + R$. Each ensemble member is then updated using a reduced-rank transformation:
\begin{equation}
    X^a_i = \bar{X}^a + X'_i\left[I - (HX'_i)^TS^{-1}_iHX'_i\right]^{\frac{1}{2}}. 
\end{equation}

\subsubsection{Ensemble Transform Kalman Filter (EnTKF)}
\label{sec:EnTKF}
The Ensemble Transform Kalman Filter (EnTKF) is a square root filter variant, designed with a more computationally efficient square root matrix than that used in the Ensemble Square Root Filter (EnSRF). We implement a version proposed initially by \cite{Bishop:2001} and subsequently refined by \cite{xuguang:2004}. Similarly to EnSRF, EnTKF performs its analysis in the ensemble subspace rather than in the whole state or observation space, as is done in stochastic EnKF.  Unlike the stochastic EnKF, which updates each ensemble member individually, EnTKF applies a single analysis update across the entire ensemble simultaneously, lowering the computational cost associated with the sequential updates. The analysis equation \eqref{eqn:SEnKF} is reformulated within the ensemble space, allowing the Kalman gain matrix (\eqref{eqn:entkf_k}) to be computed in this reduced-dimensional space rather than in the higher-dimensional observation or state spaces. The Kalman gain matrix in Equation \eqref{eqn:ensrf_mean} becomes
\begin{equation}
    K_e =  X^f\Omega(H(X'))^T R^{-1} ,
    \label{eqn:entkf_k}
\end{equation}
where $\Omega = (I + (H(X'))^TR^{-1}H(X'))^{-1}$ is the ensemble transformation matrix. The analysis update formulation for this filter is then given by:
\begin{equation}
    X^a = \bar{X}^a + \sqrt{(N_e -1)\Omega}, 
    \label{eqn:entkf}
\end{equation}
where $\bar{X}^a$ is given in Equation~\eqref{eqn:ensrf_mean} but $K_e$ replaced with Equation~\eqref{eqn:entkf_k}.

\subsubsection{Deterministic Ensemble Kalman Filter (DEnKF)}
\label{sec:DEnKF}
We implement the Deterministic Ensemble Kalman Filter (DEnKF) following the approach described by \cite{SakovDenKF}. DEnKF blends the simplicity of the Ensemble Square Root Filter (EnSRF) with features of the stochastic EnKF. Like the EnSRF, DEnKF is a deterministic filter: it yields an analyzed error covariance that does not depend on individual realizations of the observations, despite being derived from the stochastic EnKF framework. Conceptually, DEnKF can be viewed as a linear approximation to the EnSRF, particularly effective when analysis increments are small. It achieves an error covariance that asymptotically matches the theoretical value without introducing observation perturbations. In DEnKF, the Kalman analysis update is applied to each ensemble member anomaly using half the Kalman gain, $\frac{K_e}{2}$, eliminating the need for perturbed observations as shown in Equation~\eqref{eqn:denkf}. The analysis update for each ensemble member is computed via: 
\begin{equation}
    X_i^a = \bar{X}^a + X'_i - \frac{1}{2}K_eH(X'_i) ,
    \label{eqn:denkf}
\end{equation}
where $\bar{X}^a$ and $K_e$ are defined in Equation~\eqref{eqn:ensrf_mean}.

\subsection{Ensemble Initialization and Noise Modeling}
\label{sec:ens_init}
Ensemble initialization is a critical step in generating a diverse set of initial conditions that represent uncertainty in the model state. In ICESEE, this is accomplished by perturbing the initial state vector using random noise, scaled according to a user-defined ensemble spread. Noise generation is carried out either through pseudo-random fields inspired by \cite{evensen:2003}, or by using random fields produced via the \texttt{gstools} library \citep{gstools:2022} combined with Cholesky factorization. The resulting perturbed values form the initial ensemble matrix, where each column corresponds to an ensemble member and each row corresponds to a specific state variable or parameter. This structure is optimized for parallel processing during both forecast and analysis phases.

In the formulation of \citet{evensen:2003}, the state vector is augmented by a model noise vector, allowing the analysis step to jointly update covariances between observed model variables and noise components. This approach facilitates both the correction of the noise mean and a reduction in its variance.  While effective in highly dynamic and chaotic systems such as atmospheric and oceanic models--where state evolution is rapid and the system is less likely to benefit from prior ensemble realizations, this strategy may be less advantageous for ice-sheet systems, where the dominant dynamical changes occur over much longer timescales (e.g., centennial). In such slow-evolving systems, it is more plausible for the filter to extract useful information from prior realizations, and significant changes in the model noise are less likely if the initial ensemble spread is well configured. Moreover, augmenting the state vector with the model noise vector sharing the same dimensionality introduces substantial memory and computational overhead. 
In ICESEE, we implement a more tailored approach in which the model noise generated at each forecast step is written to a file and reused as input in the subsequent forecast noise generation. This method avoids unnecessary computational overhead and minimizes the risk of ensemble undersampling, which can arise from variable magnitude discrepancies--particularly since the model noise typically exhibits lower amplitude compared to state variables or parameters.  In addition, the pseudo-random field generation algorithm is modified to provide finer control over the ensemble spread during initialization and noise evolution. Specifically, we introduce a physically motivated spatial decorrelation scale, defined via the parameter {\em decorrelation\_length} as
\begin{equation}
    \text{{\em decorrelation\_length}} = \frac{\sqrt{L_x \times L_y}}{(n_x \times n_y + 1)} ,
\end{equation}
where $L_x$ and $L_y$ are the spatial extents of the domain and $n_x$, $n_y$ are the corresponding grid resolutions. This formulation ensures that ensemble variability is appropriately maintained across spatial scales and that observational sensitivity is preserved throughout the data assimilation process. Further implementation details can be found in \citet{icesee_git}.

Generating the innovation term $d_i$ in the stochastic EnKF (Equation~\eqref{eqn:_EnKF_advanced}) is important for preventing ensemble collapse and filter divergence, challenges that arise due to limited observation coverage, long-term model trends, and seasonal effects, conditions typical in ice sheet modeling. ICESEE employs two strategies to mitigate these issues that the user may choose between:

\begin{enumerate}
\item The ensemble is assumed to capture the underlying state space uncertainty already, informed by prior realizations. Ensemble perturbations ($X_i^f - \bar{X}^f$) are computed with zero mean and subsequently projected into observation space. Then the innovation term is computed using:
\begin{equation}
d_i = y^a + H(X_i^f - \bar{X}^f) .
\end{equation}
This approach reduces reliance on observation perturbations, maintains ensemble spread, and limits spurious correlations in data-sparse regimes.
\item Alternatively, ensembles of perturbations are generated using pseudo-random fields. These fields are scaled using observation standard deviations, adjusted to have zero mean, and then projected into observation space. This method helps eliminate spurious long-range spatial correlations that might otherwise result from sparse observations.
\end{enumerate}

Both methods ensure that ICESEE maintains robust ensemble spread and avoids artificial correlations, supporting stable and accurate assimilation performance over long simulation periods.

\subsection{Inflation and Localization}
\label{sec:inflation_localization}
Given the limitations of current computational resources, it is not feasible to fully represent the error covariance matrix of a high-dimensional nonlinear system using an ensemble that is substantially smaller than the model state space dimension: $N_e \ll n$. This introduces sampling errors and spurious long-range correlations due to the rank deficiency of the forecast covariance matrix, often resulting in filter divergence. To mitigate these issues, inflation and localization techniques are commonly employed to reduce such errors and improve filter performance. In many applications, including ice sheet modeling, these methods typically rely on manual tuning to identify optimal inflation and localization parameters. This process is both labor-intensive and sensitive to model configuration \citep{gillet2020assimilation, youngmin:2025}. 

For the serial implementation of the stochastic EnKF in ICESEE, we incorporate both manual and adaptive localization strategies. The adaptive approach is based on the spatially dependent Gaspari-Cohn function, utilizing ensemble correlations to compute spatially varying localization length scales by analyzing ensemble anomalies \citep{kay:localization, Evensen2022}. This is combined with either additive or multiplicative inflation. Despite these efforts, manual localization is cumbersome, and adaptive localization imposes a significant computational burden. Therefore, in this work, we primarily adopt the stochastic EnKF approach introduced in \Sec{EnKF}, along with the noise modeling strategy described in \Sec{ens_init}. These methods avoid the need for explicit localization while remaining computationally efficient and robust, as they rely on implicit use of the forecast error covariance matrix without directly computing it; see \cite{evensen:2003} for further details.

\subsection{Parameter Estimation}
\label{sec:parameter_estimation}
ICESEE can estimate model parameters by augmenting the model state vector to include both physical state variables (i.e., updated by the model) and model parameters (i.e., not updated by the model), thus constituting ``joint'' estimation. This joint vector is updated simultaneously during the analysis cycle, allowing both states and parameters to be estimated. To enable this, the parameters are artificially perturbed to introduce initial uncertainty, ensuring sufficient parameter spread (defined as the standard deviation of the ensemble parameter space). Unlike the state variables, this spread remains static during the forecast step and is only updated during the analysis phase. However, over time, joint estimation can lead to filter divergence. Specifically, the gradual collapse of parameter spread reduces the sensitivity of the filter to observations, a problem also noted by \cite{juan2013estimating, sueki2022precision}. Various mitigation strategies exist, such as localization and inflation (Section \ref{sec:inflation_localization}). Still, when the state and parameters are combined into one vector, applying localization to both state variables and parameters becomes non-trivial. As highlighted in \cite{alksoy:2006}, naive localization on the state space alone can result in a localization matrix that is not positive semi-definite, potentially leading to unphysical outcomes. However, localization is still required in the parameter space due to the existence of long-range correlations.

To address this, ICESEE adopts the dynamical model for parameters introduced by \cite{ruckstuhl2018parameter}. This approach is particularly well-suited for ice-sheet modeling, as it helps maintain parameter variability over the long response times characteristic of ice sheets. Unlike conventional Gaussian-based or inflation-based strategies, it uses a Beta distribution to describe bounded parameter evolution. A key challenge with this method is the requirement for manually specified lower and upper bounds for the Beta distribution. ICESEE overcomes this by automatically estimating these bounds using available observational data. In cases where parameters are unobserved, ICESEE adaptively retrieves lower and upper bounds from the current and previous ensemble matrices. This strategy maintains parameter spread dynamically, ensuring physically consistent updates without compromising filter stability. ICESEE applies a complementary approach by deliberately inflating the parameter subspace more than the state subspace. This is combined with the stochastic EnKF formulation described in \Sec{EnKF} and the noise modeling strategy from \Sec{ens_init}. The noise modeling approach initializes the parameter ensemble with a sufficiently large spread that is sustained throughout the assimilation cycles. This helps retain parameter sensitivity to observations and prevents ensemble collapse, further enhancing filter robustness. The experiments in this paper are performed with this complementary strategy.

In addition to joint state-parameter estimation within the EnKF framework, ICESEE can interface with external inverse modeling tools (e.g., adjoint-based methods implemented in ISSM or related optimization frameworks) to infer parameters that are weakly observed or not directly included in the assimilated state vector. In such cases, ICESEE uses the assimilated state variables as inputs to physics-based inversion procedures, enabling a hybrid assimilation-inversion workflow in which ensemble-based data assimilation and deterministic inversion are applied sequentially. In this work, ICESEE is coupled with ISSM to demonstrate this hybrid approach, where assimilated thickness and velocity fields are used to infer basal friction through ISSM’s inversion framework \citep{issm_inversion}. This combination highlights ICESEE’s flexibility in integrating ensemble-based data assimilation with established inverse modeling tools for improved parameter estimation in ice-sheet simulations.

\section{ICESEE library implementation}
\label{software_package}
The ICESEE library is developed with a modular and extensible architecture to support efficient and reproducible data assimilation for ice sheet models, and potentially other applications in the future. Its design emphasizes interoperability, flexibility, and maintainability, enabling researchers to adapt and extend the framework across diverse modeling environments. The codebase is implemented entirely in Python, selected for its user-friendly syntax, extensive scientific ecosystem, and seamless interoperability with other programming languages. Python’s capability to interface with software written in C++, Fortran, or MATLAB allows ICESEE to integrate with both legacy and modern high-performance ice sheet modeling systems.

The package structure follows a clear directory organization, with each major component accompanied by a dedicated {\em README} file describing its functionality, dependencies, and usage examples. This documentation facilitates rapid onboarding for new users and supports reproducible workflows. Detailed installation instructions, usage guidelines, and examples demonstrating coupling with various models are provided in the ICESEE GitHub Wiki~\citep{icesee_wiki}, which serves as the central hub for user and developer resources.

\subsection{Package Layout and Structure}
\begin{figure}[htbp]
  \centering
  \plotbox{\includegraphics[width=0.5\textwidth,clip=true,trim=0cm 0.5cm 0cm 1cm]{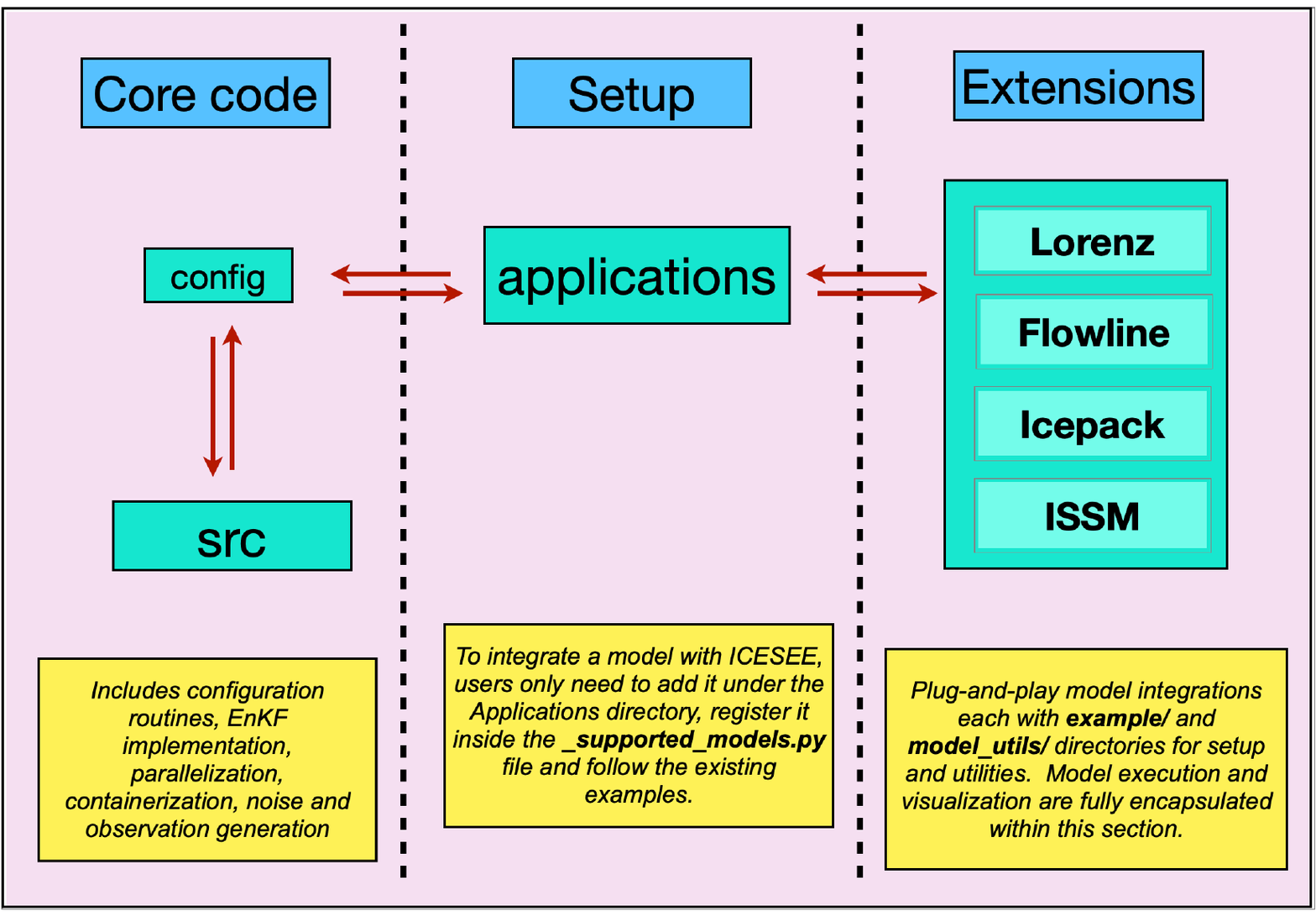}}
  \caption{High-level structure of the ICESEE codebase, showing modular components for data assimilation, model interfaces, utilities, and parallelization.}
  \label{fig:icesee-code-structure}
\end{figure}

ICESEE is organized using a systematic directory structure designed to support modular development and straightforward integration. A representative example of the actual ICESEE codebase layout is shown in \Fig{icesee-code-structure}. The framework is divided into distinct components, including model interfaces, data assimilation functions, parallelization utilities, general-purpose functions, and testing infrastructure. This modular design allows for easy integration of new models, supports both serial and parallel EnKF variants, and enables users to implement complex workflows without modifying the core assimilation logic. The bidirectional arrows in \Fig{icesee-code-structure} indicate the two-way flow of information between components. Importantly, the core ICESEE routines are kept entirely separate from the {\em applications} directory, allowing the addition of coupling capabilities for new applications without affecting the core functionality.

\begin{figure}[htbp]
  \centering
  \plotbox{\includegraphics[width=0.75\textwidth,clip=true,trim=-1cm 4.5cm 0cm 4.5cm]{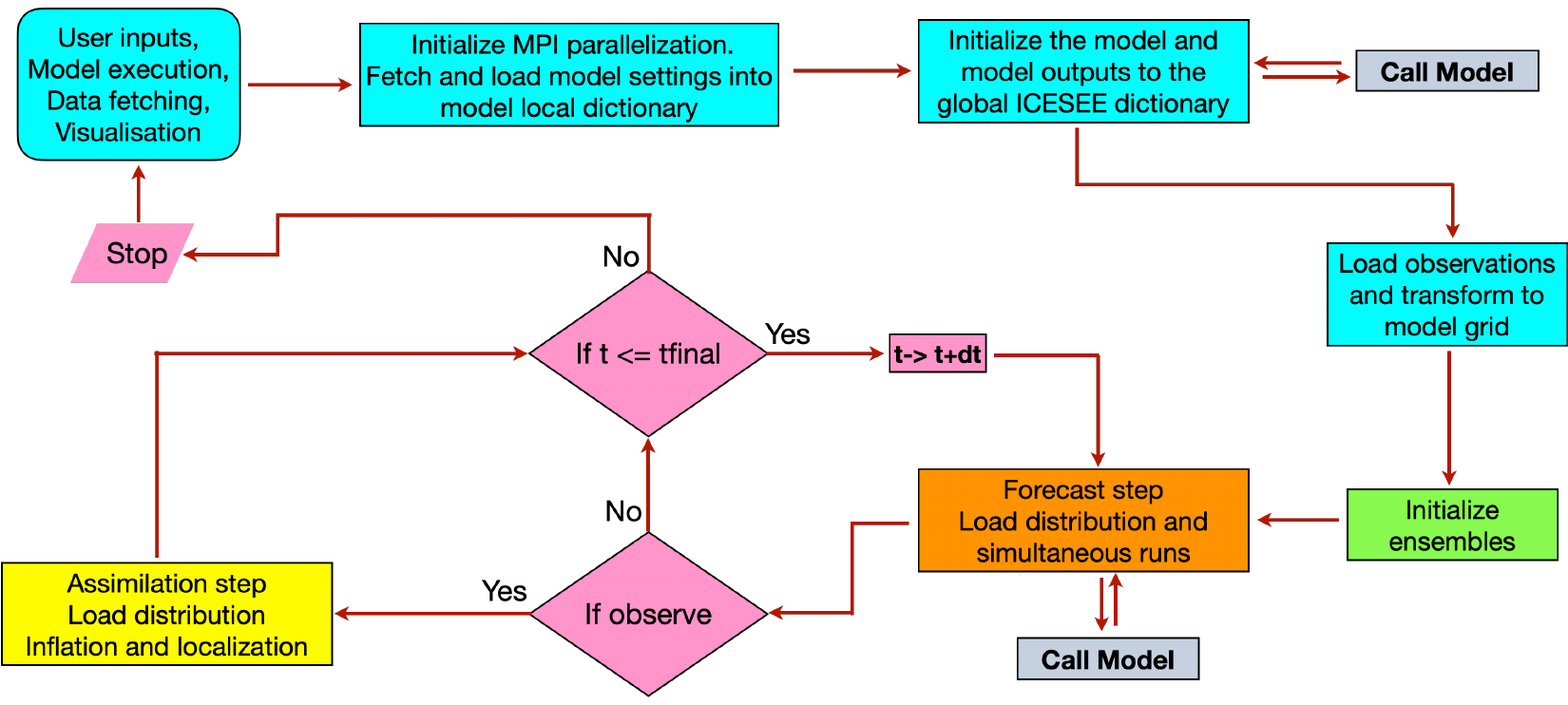}}
  \caption{Overview of the ICESEE workflow illustrating the sequence of operations from the model initializations, observation loading and transformation to the model grid, ensemble initialization, model forecasting, and assimilation. Components shown in green, orange, magenta, and yellow represent the core ICESEE routines, while those in cyan and gray correspond to model-specific extensions configurable by the user.}
  \label{fig:icesee-workflow}
\end{figure}

ICESEE also operates according to a structured workflow, illustrated in \Fig{icesee-workflow}. Each step in the workflow is executed sequentially, except for explicitly parallel processes, which run concurrently. When integrating a new application into ICESEE, the user primarily interacts with the components highlighted in cyan and gray within the workflow diagram (\Fig{icesee-workflow}). The core ICESEE routines, shown in green, orange, and magenta, are designed to be model-agnostic and can be reused across different applications. This modularity allows users to easily adapt ICESEE to their specific modeling needs without modifying the underlying data assimilation logic.

\subsection{Major Components and Routines}
This section provides an overview of the key components and routines that make up ICESEE, organized into major functional categories: data assimilation, parallel execution, utilities, configuration, application wrappers, and execution pipeline. For each category, we summarize the core routines, their directory locations, and their interactions with other parts of ICESEE and with external models.

\subsubsection{Data Assimilation Core}
\label{sec:da_core}
The data assimilation core encompasses all routines and modules related to the ensemble filtering techniques implemented in ICESEE. These routines support both serial and parallel ensemble Kalman filter (EnKF) operations, as well as auxiliary functions required for efficient data assimilation workflows. The primary components are located in the {\em src/} directory. The {\em EnKF} subdirectory contains the serial EnKF implementations written in Python, along with the Cython-based stochastic EnKF. These modules handle fundamental assimilation operations, including forecasting, analysis, ensemble perturbation, Kalman gain computation, application of the observation operator for the serial EnKF versions, and implementation of adaptive localization and inflation techniques.

The {\em parallelization} subdirectory contains modules implementing the parallel EnKF framework, including routines for parallel I/O operations, ensemble initialization, forecast and analysis steps, and creation of synthetic observations. The parallel stochastic EnKF, formulated in Equation~\eqref{eqn:_EnKF_advanced}, is implemented in both the {\em \_mpi\_analysis\_functions.py} module and the {\em EnKF\_parallel\_io.py} class, which correspond to the two parallelization strategies described in \Sec{parallelization}. The {\em run\_model\_da} directory contains the main driver script, {\em run\_models\_da.py}, which orchestrates the entire data assimilation cycle through the function {\em icesee\_model\_data\_assimilation}. This function acts as the central controller, coordinating all critical stages of the workflow: load distributions, ensemble initialization, and execution of the forecast and analysis steps. Every model integrated into ICESEE calls this function to perform its full assimilation cycle.

Each supported model, such as ISSM, Icepack, Lorenz, or the flowline model, has its own dedicated subdirectory within the {\em applications/} folder. These subdirectories include model-specific scripts for initialization, configuration, forecasting, generation of true and perturbed states (for twin experiments), and post-processing. The scripts that interface directly with the ICESEE core accept the global ICESEE dictionary (\textsl{kwargs}) as input and follow a standardized naming convention to ensure compatibility and consistency across all models. For each model integrated into ICESEE, a general data assimilation script named {\em \_modelname\_enkf.py} resides within the model’s utility subdirectory (e.g., {\em modelname\_utils/}). This general script defines the model’s EnKF routines and supports all example configurations associated with that model. However, ICESEE also allows example-specific versions of {\em \_modelname\_enkf.py} to exist within the {\em examples/} directory of a given model. These specialized scripts automatically override the general version when present, allowing for customized assimilation setups while maintaining compatibility with the ICESEE framework. This design supports both rapid prototyping of new experiments and straightforward integration of new models.

The {\em \_modelname\_enkf.py} script typically defines the following core functions:
\begin{itemize}
    \item {\em initialize\_ensemble(**kwargs)}: Constructs the ensemble matrix from the model’s true state and specified perturbation parameters, introducing the initial uncertainty for assimilation.
    \item {\em generate\_true\_state(**kwargs)}: Produces the model’s true state, used as a reference for evaluating assimilation performance in perfect model experiments. This function is called prior to the start of the assimilation cycle.
    \item {\em generate\_nurged\_state(**kwargs)}: Creates a perturbed (or “wrong”) version of the true state for perfect model experiments, representing the initial guess that the EnKF seeks to correct through assimilation.
    \item {\em forecast\_step\_single(**kwargs)}: Advances the model state forward in time for each ensemble member, generating the forecast ensemble and preparing for the subsequent analysis step.
\end{itemize}

\subsubsection{Parallel Execution}
\label{sec:parallel_execution}
This subsection describes the routines that enable parallel execution within ICESEE; a detailed explanation of the overall package parallelization strategy will be presented in \Sec{parallelization}. As illustrated in \Fig{icesee-workflow}, MPI parallelization is initialized at the start of the assimilation cycle, playing a central role in efficient data handling and in the concurrent management of ensemble members across multiple processors. These routines distribute computational workloads, handle inter-processor communication, and ensure concurrent execution of ensemble members during both the forecast and analysis phases. Most of ICESEE’s parallel workflow is coordinated by the {\em ParallelManager()} class, implemented in the {\em icesee\_mpi\_parallel\_manager} module within the {\em src/parallelization/parallel\_mpi/} directory. This class is responsible for configuring the MPI environment, assigning ensemble members to available ranks, and managing communication across the assimilation process. It provides methods for broadcasting data, gathering results, and synchronizing ensemble states across all ranks. Users typically only need to call the {\em icesee\_mpi\_init()} function to initialize the MPI environment and expose the ICESEE communicator for MPI-based workflows (see \Sec{parallelization}). This function sets up the communication infrastructure and prepares the ensemble for distributed computation. Additional MPI routines are automatically invoked within {\em icesee\_model\_data\_assimilation}, which exposes subcommunicators, ranks, sizes, and other MPI attributes to the core ICESEE routines.  

The {\em icesee\_mpi\_parallel\_manager} class also includes utilities for managing ensemble matrices, such as broadcasting ensemble states to all ranks, collecting analysis results, and computing ensemble means in parallel. Designed with flexibility and extensibility in mind, this class allows users to adapt or extend parallel execution strategies and to integrate external MPI-enabled libraries as needed.

For data management, ICESEE employs the HDF5 file format through the {\em h5py} interface~\citep{h5py_2022}, with all data stored in the {\em \_modelrun\_datasets/} directory. The HDF5 standard is particularly well-suited to high-performance computing, as it supports parallel I/O, which allows multiple ranks to read and write concurrently, making it ideal for large-scale ensemble simulations. ICESEE’s parallel I/O routines efficiently manage the reading and writing of ensemble states, observations, and other key data structures in both MPI and non-MPI environments. These routines are implemented primarily in the {\em \_parallel\_i\_o.py} module and partially in the {\em EnKF\_parallel\_io.py} class within the {\em parallelization/} directory. Together, they handle the distributed access and updating of HDF5 data structures, enabling ensemble members to concurrently read and write their output across processors while ensuring data consistency. The I/O routines are optimized to minimize communication overhead, reduce memory usage, and maximize data throughput, meeting the scalability and efficiency demands of large-scale, computationally intensive ice sheet modeling applications.

\subsubsection{Utilities}
\label{sec:utilities}
ICESEE includes a dedicated {\em utils/} directory within the {\em src/} folder, which provides a comprehensive collection of general-purpose functions supporting multiple components of the framework. This directory contains two primary modules: {\em tools.py} and {\em utils.py}. The {\em tools.py} module includes functions for automatically setting up the ICESEE Python environment, performing parallel read/write operations to HDF5 files for visualization and post-processing, and providing utilities for tracking computational performance, displaying progress bars, and summarizing runtime statistics. It also contains the adaptive indexing routine {\em icesee\_icesee\_get\_index(**kwargs)}, which generates index maps and arrays for states and parameters stored in the state vector across both MPI and non-MPI configurations. The {\em utils.py} module defines the {\em UtilsFunctions} class, which offers a broad range of reusable methods. These include generating synthetic observations at predefined observation times for perfect model experiments, computing the observation operator ($H$) and its Jacobian, and defining observation functions. These utility functions are designed with modularity and reusability in mind, allowing users to integrate them easily across different models and applications within the ICESEE framework.  Each model integrated into ICESEE may also define its own specialized utilities in a dedicated subdirectory ({\em modelname\_utils/}) within the model’s directory. These model-specific utilities can extend or override the general-purpose functions provided in the core {\em utils/} directory, enabling tailored functionality that meets the unique requirements of each model while preserving compatibility with ICESEE’s core routines.

\subsubsection{Configuration}
\label{sec:configuration}
The {\em config/} directory in the ICESEE main repository contains two primary configuration modules: {\em config\_loader.py} and {\em \_utility\_imports.py}. The {\em config\_loader.py} module is responsible for reading and parsing the {\em params.yaml} file located in each extended model directory. This YAML file defines model-specific, EnKF-related, and runtime parameters, providing a centralized and editable configuration interface that allows users to modify settings without altering the source code. It is organized into distinct sections, such as model configuration, assimilation options, and runtime controls, facilitating flexible experimentation with different assimilation strategies.

The {\em \_utility\_imports.py} module consolidates commonly used components across ICESEE. It loads parsed parameters via {\em config\_loader.py}, imports frequently used functions, dictionaries, and classes, applies general configuration settings, and retrieves command-line arguments. This structure ensures consistent access to shared parameters and utilities throughout the framework. All model-specific scripts, including {\em \_modelname\_enkf.py}, {\em \_modelname\_model.py}, {\em run\_da\_modelname.py}, and {\em \_modelname\_utils.py}, import this module to access shared dictionaries, utility functions, and configuration parameters required during data assimilation.

To interface ICESEE with supported models, the framework uses a centralized registry script, {\em supported\_models.py}, located in the {\em applications/} directory. This script maintains a list of all integrated models, their directory paths, and associated interface functions. By enforcing a uniform structure and naming convention, ICESEE ensures consistent and dynamic loading of model-specific routines during the assimilation workflow. Serving as the main configuration hub, this file allows users to specify which models to include in a given run. New models can be incorporated seamlessly by following the established registration format, promoting extensibility and simplifying integration. This centralized approach streamlines model management, enhances interoperability, and ensures compatibility with ICESEE’s core routines.

\subsubsection{Application Wrappers}
\label{sec:application_wrappers}
Because legacy ice sheet models are developed using a variety of programming languages, frameworks, and architectures, ICESEE employs dedicated application wrappers to interface effectively with these models. These wrappers are essential for automating model execution, managing data flow, and enabling seamless integration of the assimilation process across heterogeneous modeling environments.  A representative example is the coupling of ISSM with ICESEE, which requires a hybrid workflow involving communication between Python and MATLAB, as ISSM’s pre- and post-processing routines are fully implemented in MATLAB. In contrast, models such as Icepack, written entirely in Python with a Fortran computational core and a Python interface, do not require such complexity. 

To facilitate interaction with ISSM, ICESEE implements a MATLAB server (see \Sec{matlab_server}) approach: a persistent MATLAB process that listens for commands from Python and manages multiple concurrent ISSM runs. The MATLAB server implementation resides in the {\em issm\_model/issm\_utils/} subdirectory within the {\em applications/} directory. This module includes scripts for launching and terminating the server, sending command codes, handling errors (for example, terminating MATLAB processes upon failure), and other utilities that support the Python–MATLAB communication workflow. The server is designed as a general-purpose solution and can be adapted for other models requiring similar interfacing strategies.  

Since ICESEE is written in Python, implementing wrappers for external models is typically straightforward. Existing integrations demonstrate how easily ICESEE can be extended to support diverse model workflows. Each model includes a dedicated directory (for example, {\em modelname\_utils/}) located within the {\em applications/} directory, which contains the scripts required to interface that model with ICESEE.  In addition, ICESEE provides several generic wrapper utilities located in the {\em tools.py} and {\em utils.py} modules within the {\em src/utils/} directory. These include functions for mapping model states to the ICESEE main matrix, retrieving model data, and orchestrating data exchange between the model and the assimilation framework. Designed for reuse across multiple models, these tools promote consistency and significantly reduce the effort required to integrate new applications into ICESEE.

\subsubsection{Execution Pipeline} 
\label{sec:execution_pipeline}
This subsection provides an overview of the ICESEE execution pipeline, which coordinates the complete data assimilation workflow, including model setup, ensemble initialization, forecasting, and analysis, as illustrated in \Fig{icesee-workflow}. The figure also highlights how specific files and directories contribute to implementing this workflow. The cyan-shaded components in \Fig{icesee-workflow}, corresponding to tasks such as retrieving user inputs and configuration settings, initializing the model, loading the global ICESEE dictionary, and loading observations and transforming them to the model grid are implemented in the script {\em run\_da\_modelname.py}, located within each model’s example directory (for example, {\em issm\_model/examples/model-example}). This script serves as the primary entry point for executing the assimilation process for a given model, handling all key stages from setup to analysis. It imports core ICESEE routines, such as {\em icesee\_model\_data\_assimilation}, together with model-specific functions, allowing users to initiate the workflow with minimal manual configuration.  When executed, {\em run\_da\_modelname.py} loads the model parameters and runtime settings from {\em params.yaml}, configures the model, initializes the assimilation environment, and calls the {\em icesee\_model\_data\_assimilation} function. This function orchestrates the entire assimilation process, including ensemble initialization, model propagation, and analysis updates. These operations interact through the global ICESEE dictionary (\textsl{kwargs}), which serves as a centralized container for all parameters and state variables, ensuring consistent access throughout the workflow. The resulting ensemble matrices are stored in ICESEE’s primary HDF5 data structures, which serve as the central repositories for all assimilation-related information. Each forecast step updates these data structures with the propagated ensembles, which are later used in the analysis phase to assimilate available observations and compute ensemble corrections. While the execution pipeline defines the logical sequence of operations, the details of how these computations are distributed and parallelized across processors are discussed in \Sec{parallelization}.

\subsection{Coupling with ICESEE}
\label{sec:coupling_icesee}
ICESEE is designed as a flexible and extensible data assimilation framework capable of interfacing with a wide range of ice sheet models and external applications. It supports both in-memory communication and parallel file-based I/O, allowing users to select the most appropriate approach depending on the architecture and computational characteristics of the target model and the available computing resources. Integration is typically achieved through Python-based interfaces combined with model-specific APIs, enabling efficient data exchange and synchronization between ICESEE and the underlying simulation codes.

\subsubsection{In-memory Communication}
\label{sec:in_memory_communication}
In-memory communication is particularly effective for models implemented entirely in Python, such as Icepack. This approach enables direct access to model states and parameters without the overhead of file I/O, resulting in faster data exchange and shorter assimilation cycles. It is commonly realized through shared memory constructs or direct function calls, allowing real-time updates of ensemble and observation data. In-memory communication is especially well-suited for small to moderately sized ensembles and simulation size (i.e., number of elements), and for models with manageable memory requirements, as it minimizes latency and maximizes computational efficiency during the assimilation process.

\subsubsection{Parallel File-Based Input/Output}
\label{sec:file_based_communication}
For models written in languages other than Python, or for target simulations with large memory footprints and high computational demands, a file-based communication strategy provides a more scalable solution. In this approach, model states, observations, and related data are written to and read from disk using structured formats such as HDF5, which supports parallel I/O operations. File-based communication enables efficient management of large ensembles and complex models by leveraging distributed file systems and concurrent access across multiple processors. This approach is particularly advantageous in high-performance computing (HPC) environments, where memory limitations may preclude purely in-memory solutions. A representative example is the Ice Sheet System Model (ISSM), implemented in C++ with MATLAB-based pre- and post-processing. In this case, ICESEE employs file-based data exchange between Python and MATLAB, ensuring reliable integration across heterogeneous programming environments.

\subsubsection{State Index Mapping}
\label{sec:state_index_mapping}
Regardless of the communication strategy employed, ICESEE uses a dynamic routine, {\em icesee\_get\_index(**kwargs)}, to map model state indices, whether stored in memory or retrieved from HDF5 files, into the global ICESEE state matrix. This function ensures consistent access and manipulation of model data throughout the assimilation cycle. It constructs index maps and arrays that facilitate efficient retrieval and updating of state variables, enabling seamless interaction between core ICESEE routines and external model interfaces. The {\em icesee\_get\_index} routine is invoked during each data exchange phase, maintaining a consistent mapping between model-specific state variables and their corresponding entries in the global ICESEE vector. This mapping is essential for both in-memory and file-based workflows, ensuring that ensemble forecasts, updates, and analyses are applied accurately at every assimilation step.

\subsection{Package Parallelization}
\label{sec:parallelization}
ICESEE implements multiple parallelization strategies designed for the ensemble initialization, forecast, and analysis stages. These include MPI-based parallelism, Python’s {\em multiprocessing} module, and Dask. Among these, MPI serves as the primary method for large-scale stochastic EnKF simulations~\eqref{eqn:_EnKF_advanced}, providing superior scalability and computational efficiency. Python {\em multiprocessing} and Dask share similar structures to the MPI-based workflow, but are optimized for smaller-scale problems or limited computing environments. This section focuses on the MPI implementation, which is the most robust, scalable and uses widely available MPI routines for large HPC systems.

The EnKF algorithms described in \Sec{EnKF} are particularly effective for large ensemble sizes and high-dimensional state spaces typical of ice sheet models. However, managing such ensembles requires substantial computational resources, especially for processing forecasts and handling observation updates. A major advantage of EnKF methods is that the forecast step, which consumes most of the computational resources in large problems (as we show in section \ref{sec:performance}), is independent for each ensemble member. This enables {\em embarrassingly parallel} execution and excellent scalability. The analysis step, in contrast, introduces global interdependencies when computing the Kalman gain and updating the state ensemble, which can lead to communication overheads in large systems.

To address these challenges, ICESEE provides two MPI-driven parallelization modes: {\em partial parallelization} and {\em full parallelization}. In both modes, during the initialization and forecast stages, MPI sub-communicators group ensemble members into processing rounds depending on the total number of available ranks ({\em MPI ranks}) and the ensemble size ($N_e$):
\begin{itemize}
    \item \textbf{Case 1:} $N_e \ge \text{MPI ranks}$ -- The ensemble is divided into multiple rounds, each executing {\em MPI ranks} ensemble members in parallel. 
    \item \textbf{Case 2:} $N_e < \text{MPI ranks}$ -- The global communicator is split into $N_e$ sub-communicators. Each sub-communicator executes one ensemble member using $\frac{\text{MPI ranks}}{N_e}$ ranks. 
\end{itemize}

While both approaches rely on MPI sub-communicators for scalable forecast and initialization execution, they differ in how data are exchanged, stored, and distributed among ranks. The {\em partial parallelization} mode centralizes data management at the root rank, simplifying coordination but increasing memory demand--whereas the {\em full parallelization} mode distributes computation and I/O uniformly across all ranks, eliminating root-level bottlenecks. In practice, users select between the two modes based on model scale, problem size, and available resources. The partial parallelization mode is best for smaller ensembles and problem sizes or debugging runs where centralized data management simplifies coordination and reduces I/O complexity. In contrast, the full parallelization mode is preferred for large-scale experiments, where distributing computation and I/O across all ranks maximizes scalability, minimizes memory bottlenecks, and leverages high-performance parallel file systems effectively. The following subsections describe each strategy in detail.

\subsubsection{Partial Parallelization}
\label{sec:partial_parallelization}
In the partial parallelization strategy, MPI sub-communicators manage the initialization and forecast rounds, while the global root process coordinates file I/O and the analysis phase. When $N_e \geq \text{MPI ranks}$, ensemble members during the initialization and forecast stages are executed in parallel groups. Each sub-communicator root gathers local results from its ranks, aggregates them into lists, and forwards them to the global root. The global root then combines all sub-communicator outputs to reconstruct the complete ensemble array, computes the ensemble mean, and writes both outputs to disk.

During the analysis step, since the ensemble output are already available on the root rank, the global root computes $\boldsymbol{\mathcal{X}^{*}}$ and broadcasts it to all ranks. The root rank also scatters ensemble chunks to each process so that local analysis updates can be computed in parallel. The updated ensemble segments are then gathered, combined, and written back to disk for use in the subsequent forecast step.

When $N_e < \text{MPI ranks}$, the global communicator is divided into $N_e$ subgroups, each responsible for one ensemble member. Within each sub-communicator, the root process collects local state updates, which are gathered by the global root, packed, and written to disk while computing the ensemble mean after the initialization or forecast phases.

In both configurations, the analysis phase follows the same workflow: after each forecast, all ranks synchronize; the root constructs $\boldsymbol{\mathcal{X}^{*}}$; and each rank computes its local analysis update. The root rank then performs ensemble inflation, recomputes the mean following \cite{evensen:2003}, and writes both the ensemble and mean fields back to disk for the next iteration.

This approach ensures that ensemble means, perturbations, and updates are all computed and stored in parallel. Forecast and analysis data are efficiently written to HDF5 files using MPI-IO. ICESEE maintains a global dataset of dimension $(n_d, N_e, N_t)$, where $n_d$ is the state vector dimension and $N_t$ is the number of timesteps. Although this workflow provides strong scaling for both forecast and analysis phases in large ensemble data assimilation experiments, it can encounter memory constraints when the full ensemble array must reside on the root rank, particularly for very high-dimensional simulations.

\subsubsection{Full Parallelization}
\label{sec:fully_parallelization}

The full parallelization mode in ICESEE is implemented through a dedicated Python class that encapsulates all utilities required for large-scale, MPI-enabled EnKF operations. These include functions for direct parallel read and write of HDF5 dataset hyperslabs, parallel ensemble mean computation, generation of synthetic observations, construction of the observation operator, and parallel execution of the analysis update step.

During the initialization phase, the observation operator $\mathbf{H}$ and synthetic observations are computed once by the root rank and stored as Zarr files \citep{zarr} accessible to all ranks. The HDF5 ensemble datasets, with dimensions $(n_d, N_e)$, are created collectively in batches, using user-defined chunking schemes to optimize I/O performance. Each MPI rank then opens these files concurrently and retains references to its assigned dataset handles for subsequent forecast and analysis operations.

During the forecast phase, each rank reads its corresponding state slice directly from the HDF5 hyperslab, performs the forecast and noise perturbation, and writes the results back to the same dataset region in parallel. Ensemble means are computed only at observation times using collective MPI reduction operations to minimize synchronization overhead.

In the analysis phase, each rank loads its local block of the observation operator $\mathbf{H}$ and the corresponding subset of synthetic observations, computes local products ($H\bar{X}$ and $Hy$), and performs MPI {\em Allreduce} operations to assemble the global vectors across all ranks. The resulting global matrices of size $(m, N_e)$ are then used to construct $\boldsymbol{\mathcal{X}^{*}}$ concurrently on each rank. Each process reads its designated portion of the ensemble from file, applies the analysis update via matrix–vector multiplication with $\boldsymbol{\mathcal{X}^{*}}$, performs ensemble inflation, and writes the updated states back to disk.

This approach ensures full utilization of available MPI resources while minimizing communication overhead, redundant I/O, and memory consumption. By parallelizing all primary operations, from I/O to ensemble updates, ICESEE achieves near-linear scalability across large ensemble sizes and high-dimensional model states (as discussed further in section \ref{sec:performance}). This makes ICESEE particularly well-suited for advanced data assimilation experiments in ice sheet and climate modeling applications.

\subsection{Package Containerization}
\label{sec:containerization}
A recurring challenge in ice sheet modeling is deploying models on high-performance computing (HPC) platforms, where computational resources are abundant but software environments often vary widely. Containerization provides a robust solution by encapsulating all required dependencies, libraries, and configurations within a portable and reproducible environment. This capability is particularly valuable in ice sheet modeling, where different models rely on specific compiler versions, scientific libraries, and auxiliary tools that may not be uniformly available across systems.

ICESEE supports containerization for extended models that benefit from such encapsulation, most notably Icepack and ISSM. These models are complex and depend on tightly coupled software stacks that are difficult to reproduce consistently across diverse computing environments. Through containerization, ICESEE ensures that these models can be executed reliably and reproducibly, independent of the underlying hardware or operating system. For both Icepack and ISSM, ICESEE provides ready-to-use container scripts that enable users to run the models within controlled environments. Each container bundles all necessary dependencies, thereby simplifying installation, mitigating compatibility issues, and improving reproducibility. This approach also streamlines deployment across a range of platforms--from personal workstations to multi-node HPC clusters.

Currently, the Icepack container supports multi-node execution, enabling scalable parallel simulations across distributed systems. In contrast, the ISSM container is presently configured for single-node runs but can be extended to support more complex parallel configurations as needed. Both container setups are available in the {\em modelname\_utils/containers/} directory within each model’s source tree. Comprehensive build and execution instructions are provided to assist users in constructing, running, and customizing the containers according to their specific modeling and system requirements.

\section{Exemplar Results for Supported Applications}
\label{sec:supported_applications}
ICESEE currently supports four model applications: Flowline, Lorenz96, Icepack, and ISSM. In this paper, we focus on demonstrating ICESEE’s capabilities using Icepack and ISSM, as they exemplify two distinct coupling strategies and highlight the framework’s flexibility in integrating with diverse modeling architectures. Flowline and Lorenz96 applications are, respectively, a simple 1D flowline marine-terminating glacier model \citep{Robel2021s} and an implementation of the classic Lorenz model of a simple, chaotic system \citep{lorenz96}. These applications are primarily used for testing purposes.

\subsection{Icepack}
\label{subsec:icepack}
Icepack is a Python-based ice sheet model that offers a flexible and extensible framework for simulating ice dynamics. It is built on top of Firedrake \citep{firedrake}, a high-performance finite element library for solving partial differential equations (PDEs). Leveraging Firedrake’s capabilities, Icepack solves approximations of the Stokes equations that govern ice flow, making it well-suited for large-scale ice sheet simulations \citep{icepack}.

ICESEE integrates the Ensemble Kalman Filter (EnKF) with Icepack (v1.0.1) to estimate key ice sheet state variables, including ice velocity, ice thickness, and surface mass balance (SMB). The demonstration uses the synthetic ice stream example provided in the Icepack tutorials, which simulates the evolution of an ice stream containing both grounded and floating sections. For this work, we modify the example’s geometric configuration, grid resolution, and total runtime to create a customized test case that highlights the core features of both ICESEE and Icepack. This test case is designed to demonstrate how data assimilation techniques can be coupled with Icepack’s simulation engine. It illustrates ICESEE’s ability to improve predictive performance by assimilating synthetic observations into a physically realistic model. The setup showcases the full data assimilation cycle and how ICESEE facilitates real-time state estimation in a modern ice dynamics modeling environment.

\subsubsection{Icepack: Problem Setup and Model Initialization}
\label{subsec:icepack_problem_setup}
We use an elongated, fjord-like domain measuring $L_x = 5$ km in length and $L_y = 1.2$ km in width, extending from the inflow boundary to the ice front. The domain is discretized into $n_x = 12$ cells in the $x$-direction and $n_y = 8$ cells in the $y$-direction. The total simulation time is 100 years. This is meant to represent a ``small'' problem with only 96 elements and 425 degrees of freedom for each variable as a method of demonstrating some of the basic capabilities of ICESEE. All other ice sheet configuration parameters remain consistent with the original Icepack tutorial setup.

For data assimilation, we employ the parallel stochastic Ensemble Kalman Filter (EnKF) described in Equation~\eqref{eqn:_EnKF_advanced}, together with the partial parallelization strategy outlined in \Sec{partial_parallelization}. The state and parameter vector is defined as $\mathbf{X} = [h, u, v, \mathit{smb}]$, where $h$ represents ice thickness, $u$ and $v$ denote the horizontal velocity components in the $x$- and $y$-directions, respectively, and $\mathit{smb}$ corresponds to the surface mass balance. Synthetic observations are generated from the true state by adding Gaussian perturbations with standard deviations of $\sigma_{h} = 5$~m, $\sigma_{u} = 0.5$~m/s, $\sigma_{v} = 0.1$~m/s, and $\sigma_{\mathit{smb}} = 0.1$~m/yr. These observations are assimilated at selected time steps during the simulation: a total of ten observation times are used, starting at year 4 and recurring every 8 years until year 76. The ensemble size for this experiment is fixed with $N = 40$.

Initial ensemble states are perturbed using pseudo-random fields generated as described in \Sec{ens_init}, with standard deviations: $\sigma_{h} = 0.8$~m, $\sigma_{u} = 0.4$~m/s, $\sigma_{v} = 0.01$~m/s, and $\sigma_{\mathit{smb}} = 0.08$~m/yr. A decorrelation length scale of 96 m is used for all state variables except for $\mathit{smb} $, which uses a shorter length scale of 75 m to reflect greater spatial variability.

The wrong (initial guess) state is generated by modifying the surface mass balance: the accumulation rate at the inflow is reduced from 1.7 to 0.17~m/yr, and the spatial change in accumulation rate downstream is reduced from $-2.7$ to $-0.27$m/yr. Additionally, the ice thickness is perturbed by subtracting a smooth linear bump with an amplitude of 350 m, applied over the first 25\% of the domain near the inflow boundary. This bump is defined using a linearly spaced array to ensure a gradual deviation from the true state. These perturbations are simply meant to represent a common scenario where the pre-satellite glacier state and climate forcing are not well known. To compensate for under-dispersion in the ensemble, an inflation factor of 1.12 is applied to the {\em smb} updates only.

The Icepack model is initialized using the {\em initialize\_model} function, located in the {\em \_icepack\_model.py} script within the {\em icepack\_model/examples/} directory. Within this function, ICESEE establishes the MPI environment by passing an MPI communicator to Icepack through the Firedrake {\em Mesh} object. This enables Icepack to run in parallel during the forecast step, facilitating efficient computation of ensemble members. The initialization routine also reads configuration parameters from the {\em params.yaml} file, which specifies model settings, assimilation options, and runtime controls required for the simulation.

\subsubsection{Icepack: Experimental simulation}
\begin{figure*}[th]
\centering
  \includegraphics[width=\textwidth]{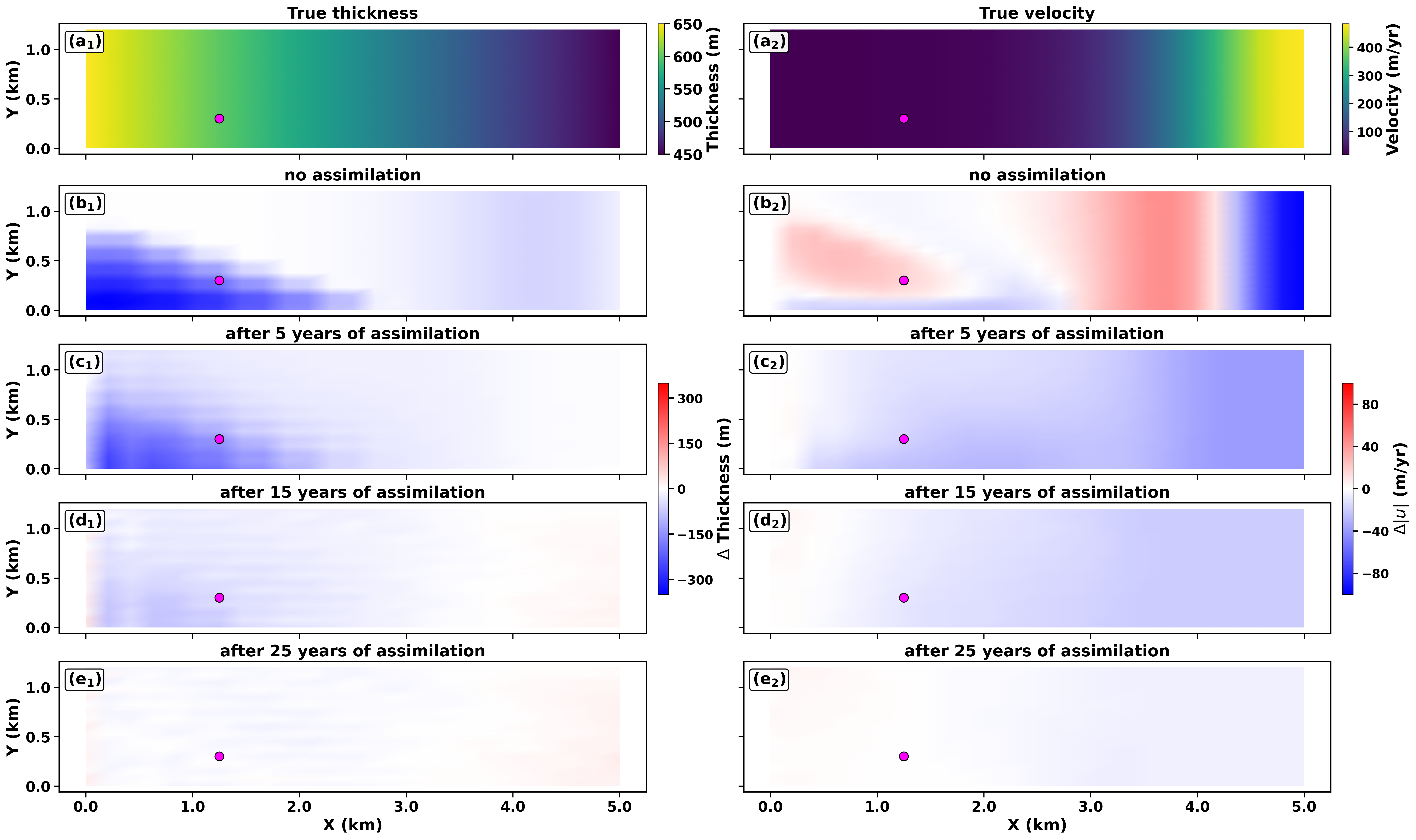}
    \caption{
    \textbf{Left side:} ($a_1$) shows the true ice thickness at time zero. 
    ($b_1$) shows the difference between the simulation without assimilation and the true thickness at time zero. Panels ($c_1$–$e_1$) show the differences between the assimilated and true thickness fields after 5, 15, and 25 years, respectively.
    \textbf{Right side:} ($a_2$) shows the true velocity magnitude at time zero. 
    ($b_2$) shows the difference between the velocity obtained without assimilation and the true velocity magnitude at time zero. Panels ($c_2$–$e_2$) show the differences between the assimilated and true velocity magnitudes after 5, 15, and 25 years, respectively. The magenta marker denotes the reference point at $(1.25~\mathrm{km},\,0.3~\mathrm{km})$, located within the perturbed (bump) region 
    of the computational domain. $\Delta |u|$ denotes the signed difference in velocity magnitude.
    }
  \label{fig:hu_diff}
\end{figure*}
Currently, ICESEE supports parallel execution with Icepack only when the number of ensemble members exceeds the number of processors. In cases where $N_{\text{ens}} < \text{MPI ranks}$, the domain is unevenly distributed across independent MPI ranks within sub-communicators. This can result in communication inefficiencies or require additional data gathering during simulation, which ICESEE does not currently support. This limitation arises from Icepack’s reliance on Firedrake, which is built on top of PETSc \citep{petsc-web-page} for parallel computing. Icepack is traditionally executed in serial mode, but in the ICESEE framework, parallelism is achieved by assigning one ensemble member per rank, allowing each rank to store and simulate the full computational domain independently. This eliminates the need for inter-rank communication during model execution and enables scalable ensemble forecasting. As a result, ICESEE can effectively perform data assimilation with Icepack on moderately large problems without incurring significant communication overhead, even when running on a single node.
\begin{figure*}[h]
\centering
  \includegraphics[width=12.3cm,clip=true,trim=0cm 0cm 0cm 0cm]{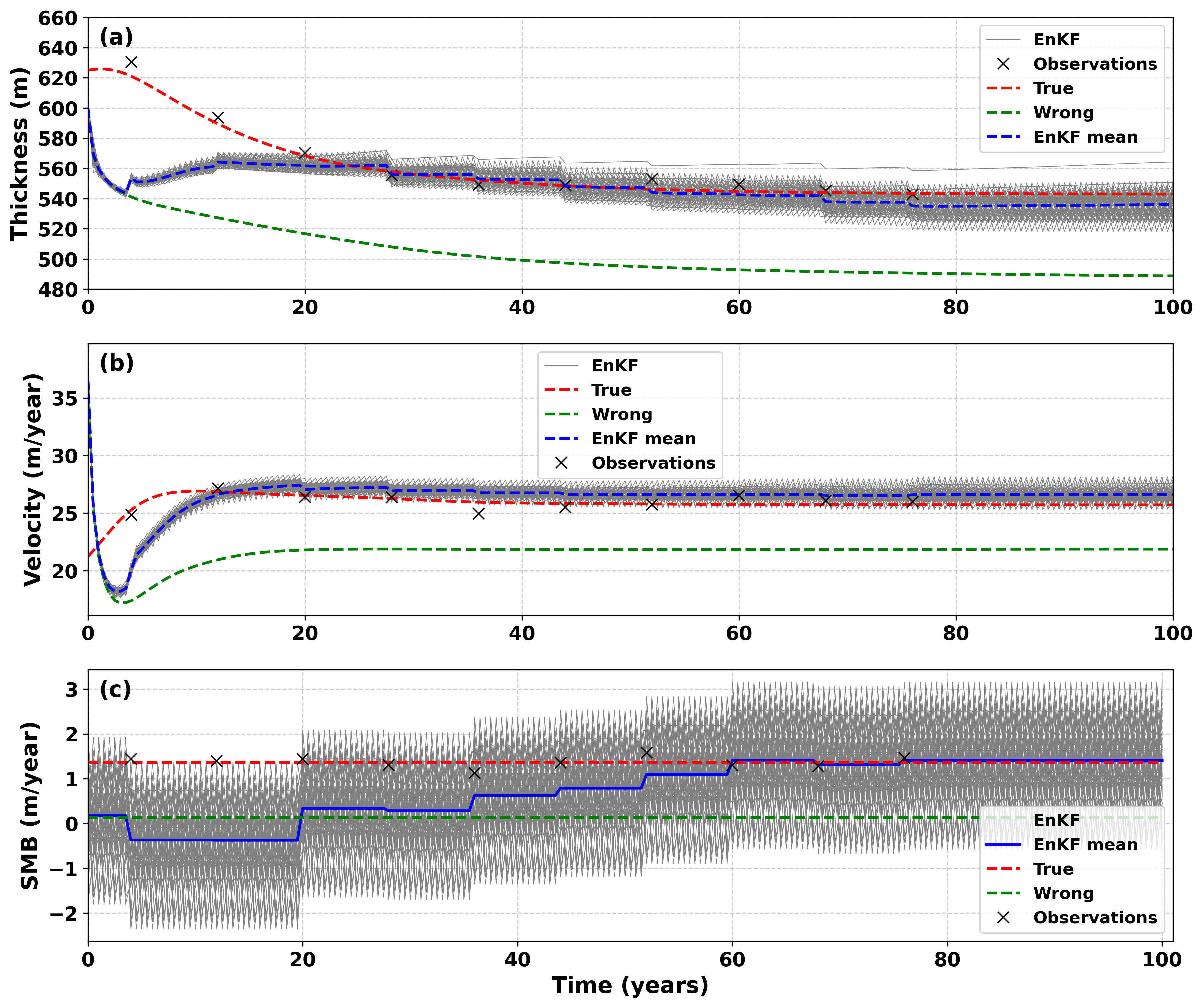}
   \caption{Temporal evolution of ice thickness (a), velocity (b),  and surface mass balance (c) at the reference point indicated by the magenta marker in \Fig{hu_diff}.}
  \label{fig:husmb_mid}
\end{figure*}

The spatial evolution of the ice thickness and velocity differences (panels ($b_1$–$e_1$) and ($b_2$–$e_2$) in \Fig{hu_diff}, respectively) demonstrates the ability of ICESEE to recover the true simulation by 25 years, using only four assimilation cycles (see \Fig{husmb_mid}). The magnitude of the thickness difference decreases substantially, from approximately 350~m to values close to zero. Similarly, the velocity difference is reduced from about 85~m\,yr$^{-1}$ to values near zero. These results highlight the robustness of the ICESEE-Icepack coupling in correcting the initially biased state.

The joint estimation capability of ICESEE is illustrated in \Fig{husmb_mid}. At the reference point, the ensemble mean for ice thickness converges toward the true state after around 20~years, while the velocity field aligns with the true value within the first 10~years. This demonstrates that the filter rapidly corrects the dynamical variables once sufficient observational information becomes available. The ensemble spread around the mean remains nearly constant throughout the simulation, indicating stable uncertainty representation and limited ensemble collapse.

The temporal evolution of the surface mass balance (\Fig{husmb_mid} (c)) reveals greater sensitivity to the assimilated observations, with variability persisting until the final assimilation stage. After about 60~years, the filter successfully corrects $\mathit{smb}$ toward the true value, showing that ICESEE can recover slowly evolving or weakly constrained parameters that are not directly observed. In general, these results confirm that ICESEE effectively constrains both fast and slow components of the coupled ice-sheet system through joint state–parameter estimation, achieving robust and stable convergence.

\subsection{Ice-sheet and Sea-level System Model (ISSM)}
\label{sec:issm}
ISSM is a comprehensive ice sheet modeling framework that captures a wide range of physical processes governing ice dynamics, including ice flow, thermodynamics, and interactions with the ocean and atmosphere. ISSM is designed to simulate ice sheet evolution over long timescales, providing critical insights into their contributions to sea-level rise and their sensitivity to climate change. The model is implemented in C++ and features a MATLAB-based interface for pre- and post-processing, enabling flexible configuration and analysis workflows \citep{issm}.

In this work, we demonstrate ICESEE’s capabilities by coupling it with ISSM using the Ensemble Kalman Filter (EnKF) for data assimilation. This integration illustrates how ICESEE can improve ISSM's predictive performance by incorporating synthetic observations to constrain and forecast key variables, including grounding line position, thickness, velocity, and bed topography.

The ISSM application underscores ICESEE’s adaptability in interfacing with complex modeling systems that require specialized integration techniques. In particular, we use a MATLAB server-based communication approach to enable seamless interaction between ICESEE (Python) and ISSM (MATLAB/C++), demonstrating the framework’s ability to bridge language and platform boundaries for advanced data assimilation tasks.

\subsubsection{MATLAB Server}
\label{sec:matlab_server}

The MATLAB server class ({\em MatlabServer}) is a key component of the ICESEE–ISSM integration, enabling efficient and automated communication between Python and MATLAB. It allows ICESEE to execute ISSM commands and retrieve model outputs without requiring users to run MATLAB scripts manually. The server continuously listens for commands issued by the Python environment, executes them in MATLAB, and returns the results to ICESEE. To ensure non-blocking communication, MATLAB output is handled using multithreading, which enables the real-time capture and display of MATLAB console output on the Python side without interrupting ICESEE’s main processes. Based on the number of ensemble members, the server can manage many concurrent ISSM runs by launching separate MATLAB instances equal to the number of ensemble members or requested processes.

We implement a two-level parallelism strategy: each MATLAB instance is initialized with a specific number of internal ranks, while ICESEE handles the distribution of ensemble members across those instances. This design achieves fully parallelized execution across both ISSM and ICESEE. The server is implemented in the {\em mat2py\_utils.py} script located in the {\em issm\_model/issm\_utils/matlab2python} directory. This module contains all supporting routines for managing server processes, including launching and terminating MATLAB instances, sending command codes, handling exceptions (e.g., terminating stalled MATLAB processes), and coordinating communication.

Communication between MATLAB and Python is implemented using file-based parallel I/O, as described in \Sec{file_based_communication}. The server is designed to be robust and flexible, with configurable behavior to support different use cases. Although tailored for ISSM, the server is general-purpose and can be adapted to other models that require a similar Python–MATLAB coupling strategy.

\subsubsection{ISSM: Synthetic Twin Configuration}
\label{sec:issm_problem_setup}
We consider a modified version of the ISMIP benchmark problem originally described by \cite{gillet2020assimilation} and later adapted by \cite{youngmin:2025} to simulate the long-term evolution of a marine-terminating glacier in 3D. The model domain is a rectangular region of size $L_x = 640~\text{km}$ and $L_y = 80~\text{km}$, representing an idealized marine-terminating ice-sheet geometry. All simulations are performed using ISSM version 4.24 and are coupled to ICESEE for ensemble-based data assimilation. A reference (truth) simulation is constructed following the general configuration of \cite{youngmin:2025}, with targeted modifications to the basal and bed topography as well as mesh discretization to ensure a steady-state solution for the present case study. In particular, we adjust the bed ($b$) geometry coefficients controlling the characteristic sidewall width ($f_c$), trough half-width ($w_c$), and the trough depth relative to the sidewalls ($d_c$). These parameters are modified from $f_c = 4000$ to $3000~m$, $w_c = 24000$ to $23000~m$, and $d_c = 500$ to $400~m$, respectively.  We further modify the surface and base parameterizations: the surface elevation is changed from $\max(b_x(x) + b_y(0) + 100, 10)$ to $\max(b_x(x) + b_y(0) + 300, 20)$, while the base is adjusted from $\max(b, -90)$ to $\max(b, -280)$. The computational domain is discretized using $n_x = 222$ grid cells in the $x$-direction and $n_y = 56$ cells in the $y$-direction (corresponding to average horizontal spacings of approximately $2.9$ km in $x$ and $1.4$ km in $y$, with a non-uniform mesh featuring local element sizes ranging from about $500$ m in refined floating-ice regions to $10$ km in coarser grounded-ice regions).  After parameterization, exponential noise is added to the bed topography, and a constant offset of $284$~m is subtracted from every node in the domain. Following these modifications, we perform a $25{,}000$-year spin-up simulation to obtain a steady-state reference configuration. The surface mass balance ($smb$) is maintained as in \cite{youngmin:2025}, while the basal melt rate is reduced from $200$ to $150~\text{m}\,\text{yr}^{-1}$. The model is then run forward for an additional 100 years to produce the final ``true'' reference simulation used in the assimilation experiments.

The ICESEE state vector is defined as
\[
\mathbf{X} = [h,\, \mathit{s},\, u,\, v,\, \mathit{b},\, \mathit{c}],
\]
where $h$, $s$, $u$, and $v$ denote the state variables (ice thickness, surface elevation, and horizontal velocities, respectively), and $b$ and $c$ denote the state parameters (bed topography and basal friction coefficient) that are estimated jointly within the assimilation framework. The corresponding decorrelation length scales used for pseudo-random field generation during the forecast step are $[88,\,85,\,90,\,85,\,88,\,100]$~m for $[h,\,s,\,u,\,v,\,b,\,c]$, respectively.

Following \cite{youngmin:2025}, we use {\em GSTools} to generate spatially correlated random fields: a Gaussian random field for the basal friction coefficient and an exponential kriging field for the bed topography conditioned on bed observations. The friction variogram parameters are modified to better match the adjusted bed geometry, with the sill reduced from $90{,}000$ to $10{,}000$ and the range reduced from $5{,}000$ to $500~m$. Ensemble initialization is performed by nudging the mean friction field ($2500~Pa~m^{-1/3}yr^{1/3}$) using the generated friction noise, while the bed, base, and surface fields are perturbed using the bed noise for each ensemble member. In addition, a uniform offset of $50$~m is subtracted from the bed and base elevations, and $25$~m from the surface elevation, across the entire domain. These perturbed fields serve as ``wrong'' initial conditions for the ISSM transient solver, yielding the ensemble initial states $\mathbf{X}$, and are selected to be representative of the scale of errors used in initial conditions for ice sheet simulations.

Synthetic observations of surface elevation and the $x$- and $y$-components of velocity are generated from the true state by perturbing it with additive Gaussian noise. The prescribed standard deviations are $\sigma_s = 2.5~\mathrm{m}$ for surface elevation, $\sigma_u = 3.0~\mathrm{m~yr^{-1}}$ and $\sigma_v = 2.0~\mathrm{m~yr^{-1}}$ for the horizontal velocity components, and $\sigma_b = 2.0~\mathrm{m}$ for bed topography. Surface elevation and velocity observations are assumed to be available everywhere over the domain (similar to a region where satellite-based altimetry and SAR data might be available), whereas bed topography observations are restricted to a two-dimensional sampling mask consisting of aircraft-based radar flight lines oriented along the $x$-direction, with adjacent lines spaced every $4000~\mathrm{m}$, to mimic realistic airborne surveys. Bed topography is observed once at year 2, while surface elevation and velocity observations are assimilated every two years through year 24 of the simulations. Ice thickness $h$ is not observed directly; instead, it is inferred during assimilation through cross-covariances between model state variables and parameters.

Basal friction is not observed directly but is inferred at each assimilation cycle from the velocity ensemble in combination with synthetic velocity observations, effective pressure (computed from the assimilated ice thickness), and grounding line evolution (derived from the assimilated thickness and bed topography). This inference is achieved through the coupling of traditional ice sheet velocity inversion methods \citep{macayeal1989large,issm_inversion} within ISSM and the ICESEE data assimilation framework, with Tikhonov regularization applied to stabilize the inversion. The basal friction coefficient is constrained to lie between $2000$ and $4000~\mathrm{Pa ~m^{-1/3} yr^{1/3}}$, applied uniformly across the domain. The absolute and relative velocity misfit weights are both set to 1, while the regularization weight is fixed at $10^{-13}$. Velocity inversion for basal friction coefficients occurs whenever velocity data is available and replaces the prior basal friction estimate.

During ICESEE data assimilation runs, we use a time step of $0.2$ years and are integrated forward for a total of 50 years, which is sufficient to capture the effects of interest in this study. The ensemble size for this experiment is again fixed with $N = 40$. Because the bed is observed only once, we employ an adaptive under-relaxation strategy at the snapshot time,
\begin{equation}
b^{\text{new}} = b^{\text{prior}} + \alpha(t)\left(b^{\text{assimilated}} - b^{\text{prior}}\right).
\end{equation}
The relaxation parameter $\alpha(t)$ is constructed from both deterministic and stochastic components,
\begin{equation}
\alpha(t) = (\eta + \beta_t)\Delta t + \sqrt{\Delta t}\,\sigma_b\,\rho,
\end{equation}
where $\eta$ is a baseline adjustment coefficient, $\beta_t$ is an ensemble-derived bias term,  and $\rho$ is a stochastic parameter (see \cite{evensen:2003} for its calculation). This structure allows the update magnitude to adapt to ensemble statistics while incorporating controlled variability.  The relaxation parameter is capped at $\alpha(t) \le 0.5$, enforcing a contractive update that limits the applied correction to at most $50\%$ of the inferred bed increment. This controlled update mitigates filter over-adjustment, suppresses spurious high-frequency noise, and stabilizes the ensemble in the absence of subsequent bed observations.

\subsubsection{ISSM: Synthetic Twin Results}
\label{sec:issm_experimentalsimulations}
\begin{figure*}[h]
\centering
\begin{subfigure}[t]{0.495\textwidth}
  \centering
  \includegraphics[width=\linewidth]{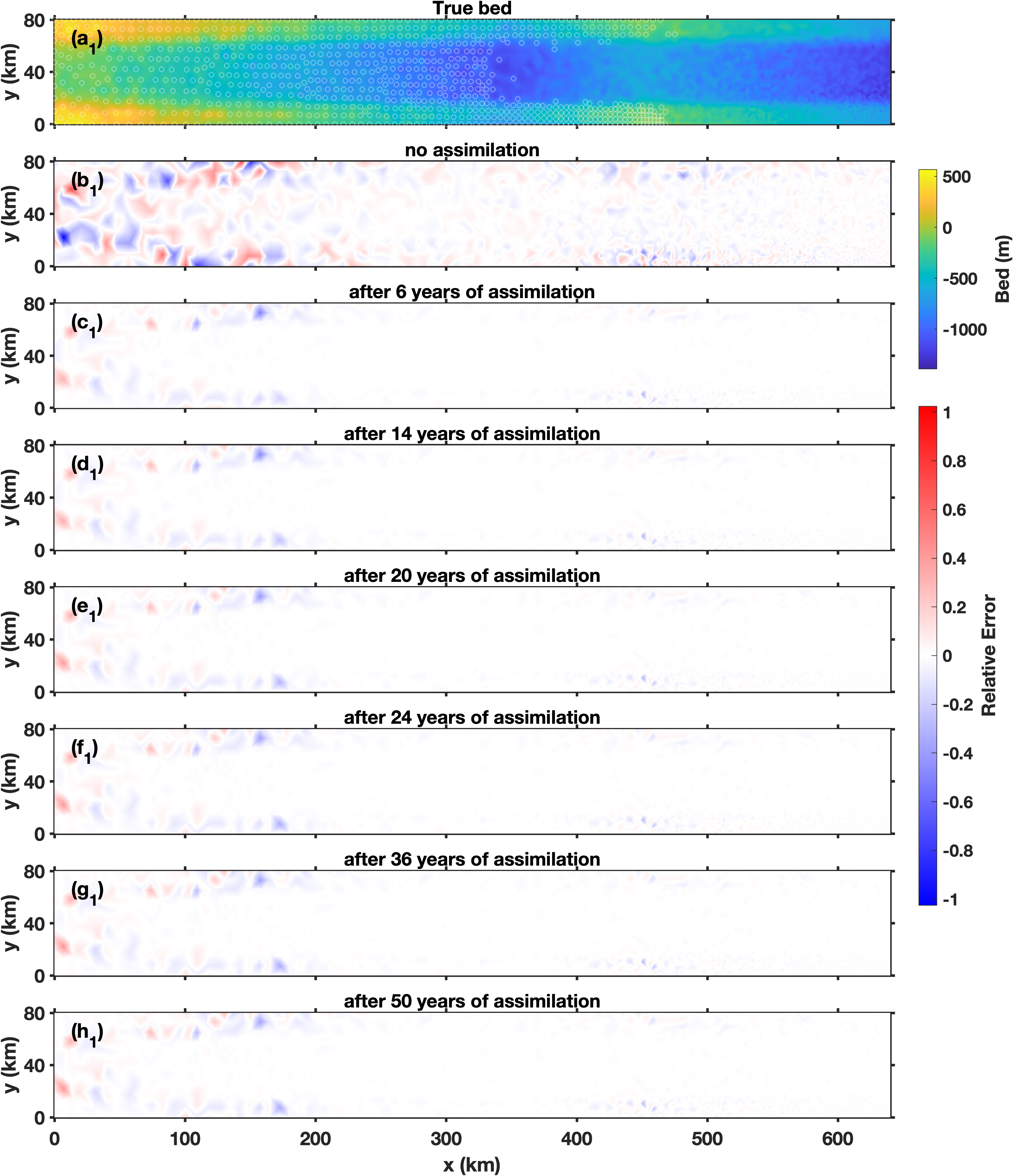}
  \label{fig:beddiff}
\end{subfigure}
\hfill
\begin{subfigure}[t]{0.48\textwidth}
  \centering
  \includegraphics[width=\linewidth]{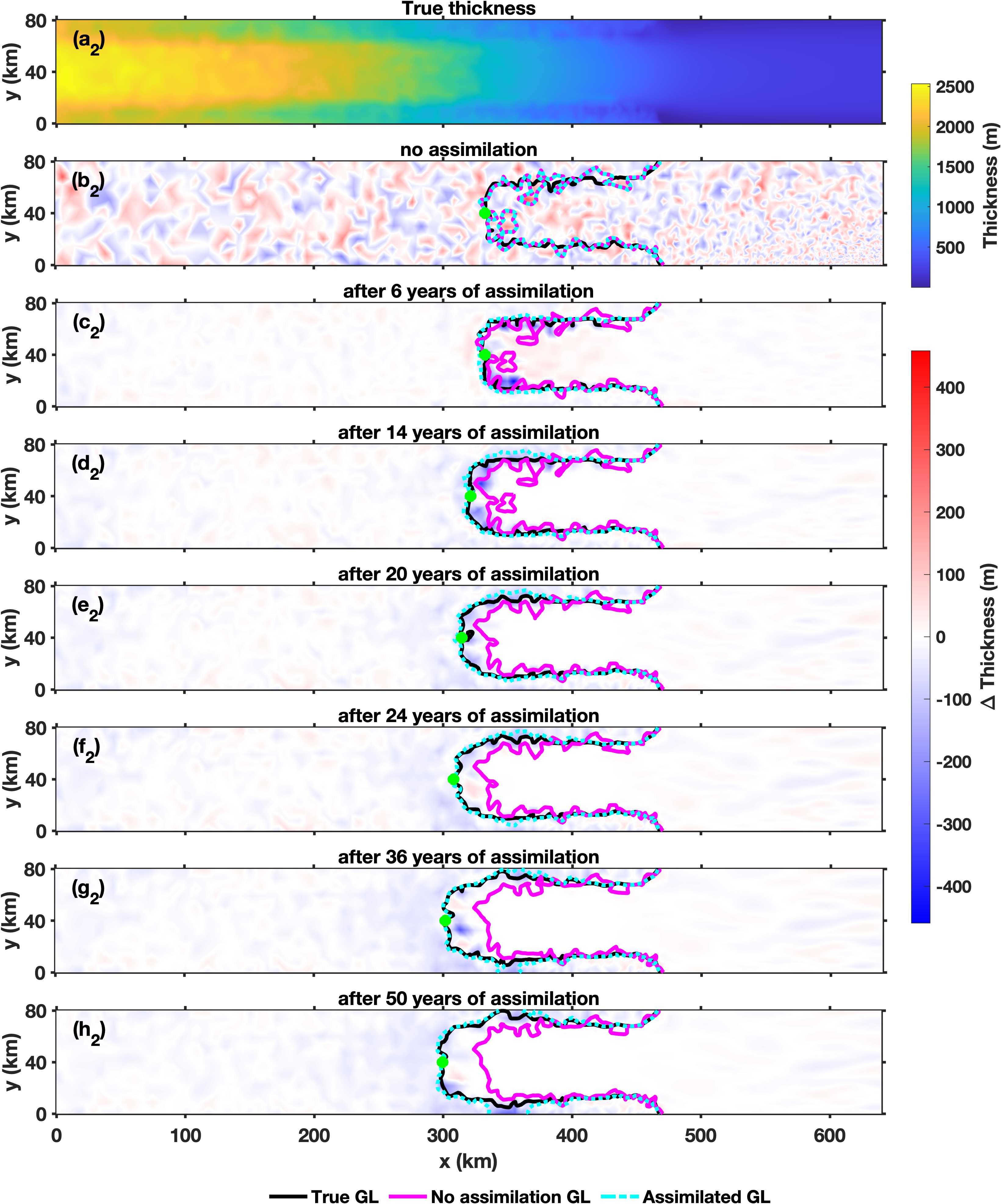}
  \label{fig:thicknessgl}
\end{subfigure}
\vspace{-0.68cm}%
\caption{
\textbf{Left column:} Reference bed topography ($\text{a}_1$), bed elevation error of the unassimilated simulation relative to the true state ($\text{b}_1$), and bed elevation error of the assimilated simulation relative to the true state at selected times ($\text{c}_1$–$\text{h}_1$). The small white circles overlaid on panel $\text{a}_1$ mark the locations of bed observations assimilated at year~2.
\textbf{Right column:} Reference ice thickness at the initial time ($\text{a}_2$), ice thickness difference between the unassimilated and true simulations ($\text{b}_2$), and the temporal evolution of the ice thickness difference between the assimilated and true states ($\text{c}_2$–$\text{h}_2$). The black solid, magenta solid, and cyan dashed curves denote the grounding line (GL) positions for the true, unassimilated, and assimilated states, respectively. The green solid dot on the true grounding line denotes the point at which the centerline intersects the grounding line.
}
\label{fig:geometry_gl_bed}
\end{figure*}

\begin{figure*}[t]
\centering
\begin{subfigure}[t]{0.49\textwidth}
  \centering
  \includegraphics[width=\linewidth]{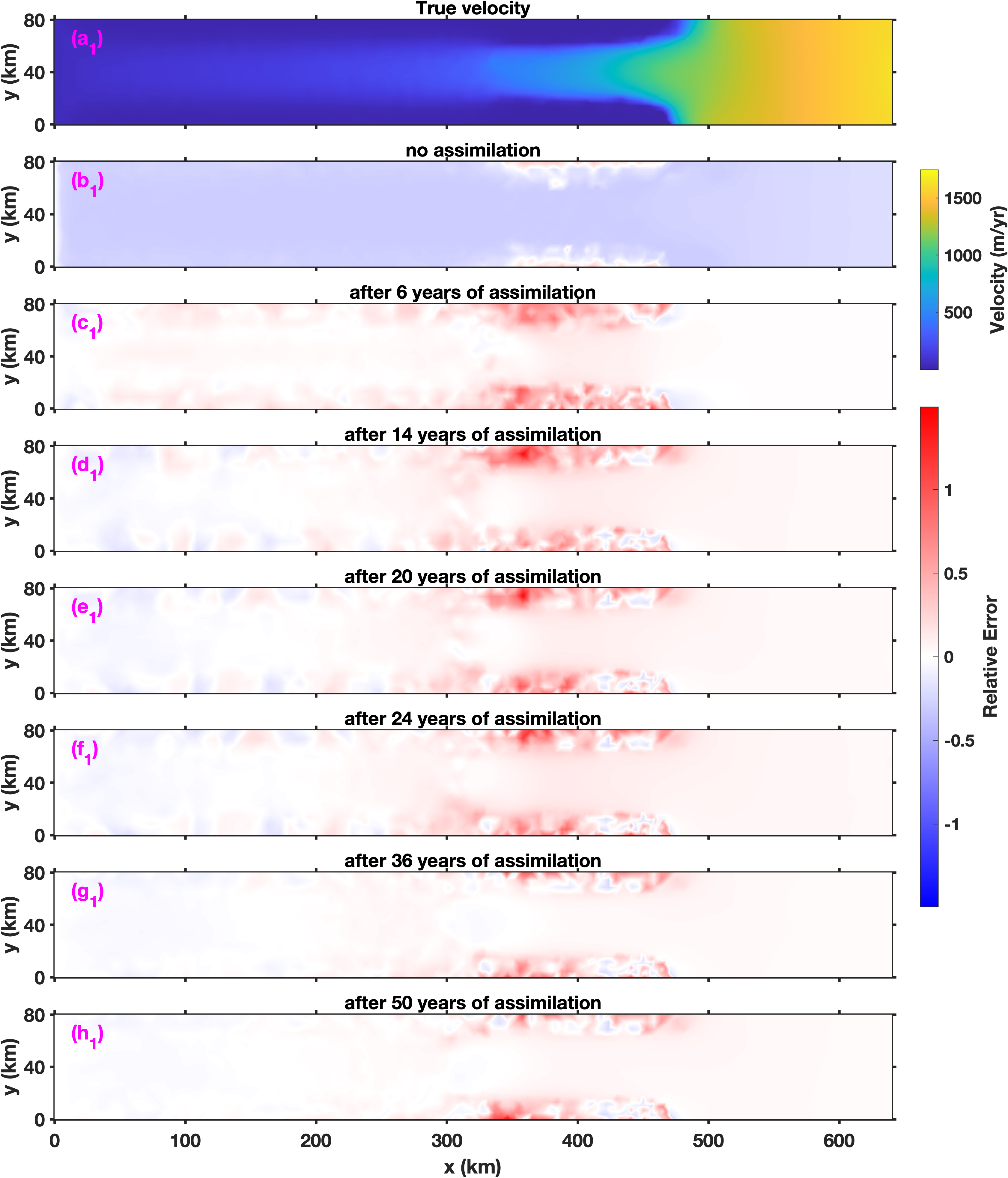}
  \label{fig:veldif}
\end{subfigure}
\hfill
\begin{subfigure}[t]{0.49\textwidth}
  \centering
  \includegraphics[width=\linewidth]{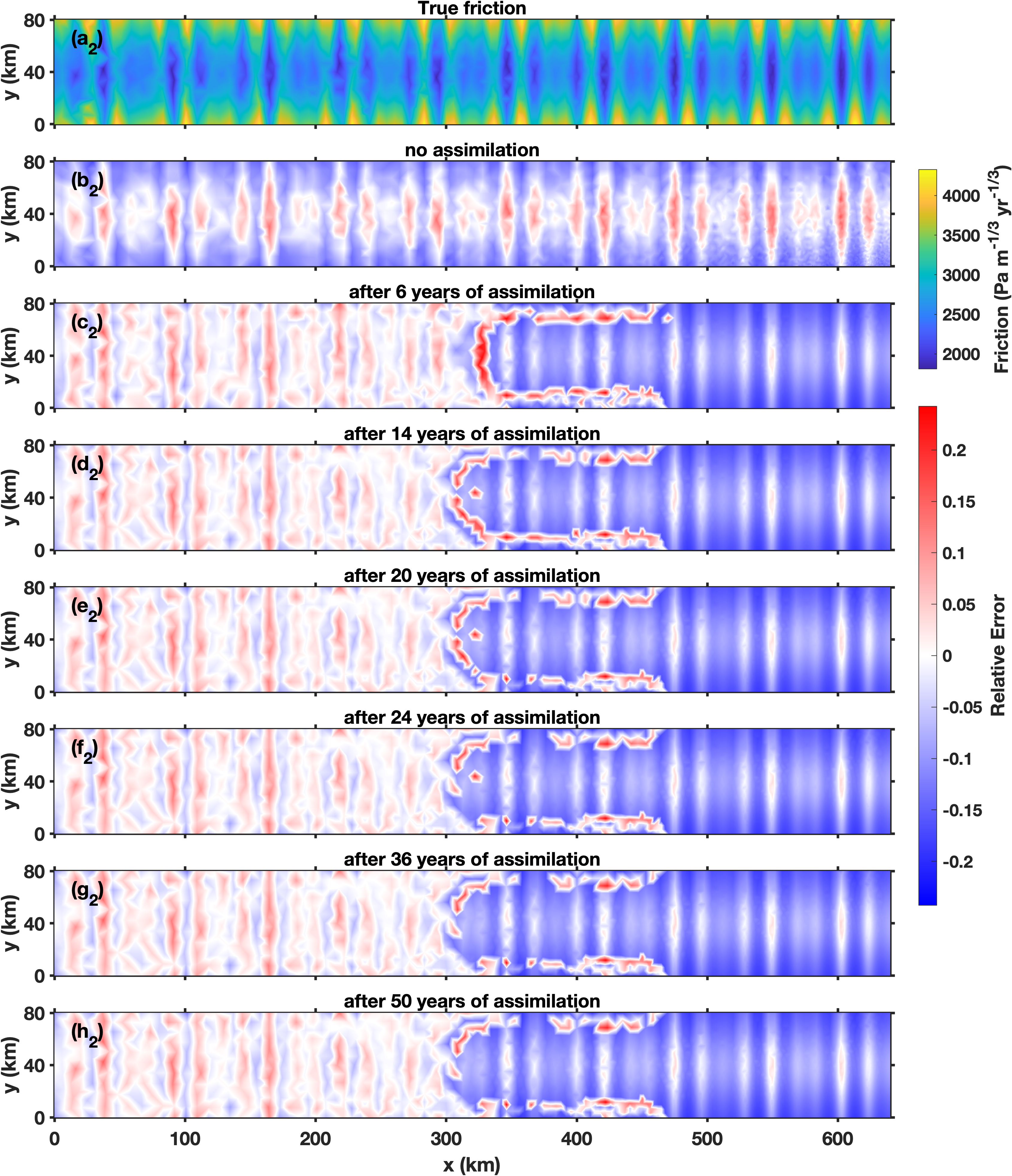}
  \label{fig:fcoef}
\end{subfigure}
\vspace{-0.68cm}%
\caption{\textbf{Left side:} Reference velocity field and unassimilated velocity error relative to the true state at time 0 ($\text{a}_1$, $\text{b}_1$), respectively, followed by assimilated velocity error relative to the true velocity at selected times ($\text{c}_1$–$\text{h}_1$). \textbf{Right side:} Reference basal friction coefficient ($\text{a}_2$), the basal friction coefficient difference between the unassimilated and true states ($\text{b}_2$), and the difference between the assimilated and true basal friction coefficients at successive assimilation cycles ($\text{c}_2$–$\text{h}_2$). }
\label{fig:friction_vel}
\end{figure*}

\begin{figure*}[t]
\centering
  \includegraphics[width=0.805\textwidth]{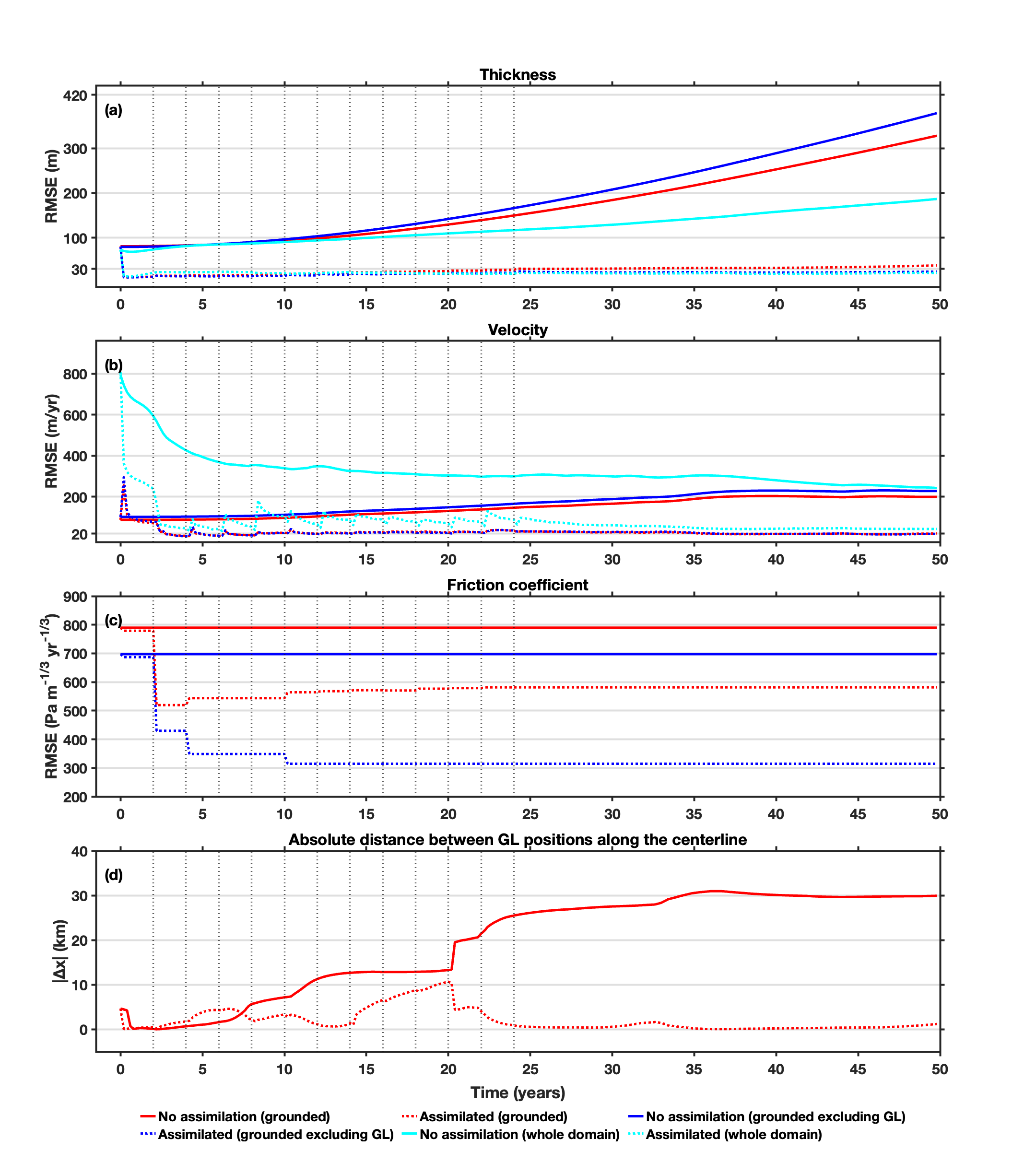}
   \caption{Temporal evolution of the root-mean-square error (RMSE) of ice thickness, velocity, and basal friction coefficient (a–c), and the absolute difference between grounding line positions for the assimilated and true states, as well as for the unassimilated and true states, measured along the horizontal centerline passing through the green solid dot on the true grounding line in \Fig{geometry_gl_bed} (right side) (d). Thin gray dotted vertical lines indicate assimilation times. Dotted red, blue, and cyan curves correspond to assimilated runs evaluated over grounded ice, grounded ice excluding the region upstream of the grounding line, and the whole domain, respectively; solid curves show the corresponding unassimilated cases.}
  \label{fig:hucgl}
\end{figure*}

Here we assess the ability of the ICESEE data assimilation framework to correct the erroneous model state $\mathbf{X}$ using the coupling between ICESEE and ISSM described above. We summarize the results in Figures ~\ref{fig:geometry_gl_bed}, \ref{fig:friction_vel}, and \ref{fig:hucgl}, which demonstrate that the coupling works as intended and that the data assimilation method implemented in ICESEE successfully corrects the model state towards the ``true'' state. On the left side of \Fig{geometry_gl_bed}, the bed topography is observed directly only once at year~2. Subsequent corrections are achieved through indirect information learned by the EnKF cross-covariances between model variables and parameters. In general, relative errors within 100 km of the grounding line, and in the central trough of the glacier are less then 10\% after the first few assimilation cycles, though relative errors remain locally high upstream and in the slow-flowing margins. Other studies using data assimilation to estimate bed topography find similar patterns of better estimates of bed properties in fast-flowing areas \citep{gillet2020assimilation, brinkerhoff2025, youngmin:2025}. However, the overall reduction in the relative bed-elevation error at selected time steps demonstrates the effectiveness of this indirect correction even with limited observations.

The right side of \Fig{geometry_gl_bed} ($\text{b}_2$–$\text{h}_2$) shows the difference in temporal evolution of ice thickness and grounding line position in simulations with and without data assimilation.  With assimilation implemented through ICESEE, errors in ice thickness are substantially reduced and observed grounding line retreat tracks very closely to the ``true'' state even though grounding line position is not explicitly assimilated in these simulations. Further evidence of the improvement in thickness and grounding line position is provided in \Fig{hucgl}. Panels~(a) and~(d) quantify these corrections using the root-mean-square error (RMSE) of ice thickness and the absolute difference between grounding line positions for simulations with the ``true'' state, assimilation (dashed lines), and no assimilation (solid lines), measured along the horizontal centerline passing through the green solid dots shown in the left side of \Fig{geometry_gl_bed}. The thickness RMSE (\Fig{hucgl}a) increases gradually for the case without assimilation (solid lines), while the case with assimilation (dashed lines) exhibits a rapid reduction in  RMSE during the first year, decreasing from roughly 80 m to below 30 m, after which the error remains approximately stable for the rest of the simulation since the dominant initial biases are corrected. 
Similarly, \Fig{hucgl} (d) shows that the grounding line position in the simulation with assimilation (solid line) progressively diverges from the true state accumulating more than 30 km of error within a few decades. In the case with assimilation (dashed line), grounding line position converges toward the true grounding line and ends with error less than the grid spacing. This behavior is consistent with the grounding line evolution shown in \Fig{geometry_gl_bed} and confirms that ICESEE data assimilation effectively corrects bed errors and improves grounding line evolution without explicitly assimilating observed grounding line positions. The bed corrections remain relatively small, as expected given the limited availability of bed observations following the initial set of observations. 

ICESEE is capable of updating estimates of unobserved parameters, such as the basal friction coefficient, through more traditional inverse methods (e.g., the deterministic adjoint-based inversion framework implemented in ISSM). Because viscous ice sheet flow is non-inertial, the velocity solvers in all modern ice sheet models, in which the basal friction coefficients appear, are completely diagnostic. Therefore, in sequential data assimilation methods such as those implemented in ICESEE, there may be a value to updating basal friction through purely inverse methods. ICESEE is configured to allow the user to choose whether to update basal friction fields via the EnKF or via traditional inversion. When the inversion option is chosen, during the forecast step, the full model state vector $\mathbf{X}$ is evolved forward in time. However, at each assimilation step, $\mathbf{X}$ is reduced to exclude the friction coefficient, and the assimilation for other variables and parameters is performed using cross-covariances between thickness, surface elevation, velocity, and bed topography, combined with velocity and surface observations available everywhere over the domain and restricted, constrained bed observations available only once at year~2. Following each assimilation step, ICESEE uses the assimilated thickness and velocity fields, together with velocity observations and the prior (initial) friction field, to infer the basal friction coefficient via ISSM’s inversion framework \citep{issm_inversion} and associated cost functions. The inferred friction field is then appended back to the model state $\mathbf{X}$ for the next forecast. 

The inferred friction differences between the true and assimilated fields at selected assimilation times are shown in the right side of \Fig{friction_vel} ($\text{c}_{2}$–$\text{f}_2$). After year~24, no further updates to the friction coefficient are observed--see ($\text{g}_{2}$–$\text{h}_2$), as this corresponds to the final observation time. The ``true'' velocity at time~0 ($\text{a}_{1}$), together with the velocity error relative to the true state for the cases without assimilation ($\text{b}_1$) and with assimilation ($\text{c}_{1}$–$\text{h}_1$) cases, are shown in \Fig{friction_vel}. The results show a large improvement in the assimilated case, with basal friction errors reduced to below 5\% in grounded regions and velocity errors reduced to approximately 10\% upstream and downstream of the grounding line following the first inversion at year 2, and then continued more gradual improvement as the inversion incorporates other state variables, which are also gradually improving with subsequent assimilation cycles. Friction is set to zero beneath floating ice where large velocities would otherwise bias the inversion. Consequently, the assimilation more effectively constrains friction in grounded regions, leading to improved velocity agreement with the true state at each assimilation cycle. These findings are further supported by the velocity and friction RMSE diagnostics shown in \Fig{hucgl} (b) and (c), respectively. The velocity RMSE exhibits sharp peaks at assimilation times, reflecting the diagnostic nature of the velocity solver combined with the sparsity of the velocity observations. 

ICESEE performs more effectively in grounded ice regions upstream of the grounding line than within or downstream of the grounding line zone, as evidenced by ice thickness (\Fig{geometry_gl_bed}, panels $\text{b}_2$–$\text{h}_2$, and \Fig{hucgl}a), velocity (\Fig{friction_vel}, panels $\text{b}_1$–$\text{h}_1$, and \Fig{hucgl}b), and basal friction (\Fig{friction_vel}, panels $\text{b}_2$–$\text{h}_2$, and \Fig{hucgl}c). For example, \Fig{hucgl} shows that the assimilated RMSE of basal friction for grounded ice excluding the grounding line (blue dashed curve) is consistently lower than both its unassimilated counterpart and the assimilated RMSE computed over grounded ice including the grounding line. Similar behavior is observed for velocity and thickness, as well as for the whole-domain assimilated RMSE relative to the unassimilated case for velocity and thickness. This behavior reflects the fact that the filter learns most effectively from variables and parameters that are well constrained by both observations and model physics, conditions that are predominantly satisfied in grounded ice regions. In contrast, the grounding line zone is characterized by strong nonlinearities and increased model uncertainty, which limit the filter’s effectiveness. Nevertheless, these results demonstrate that ICESEE successfully exploits cross-variable covariances to propagate bed information and improve ice-sheet geometry and dynamics through its coupling with ISSM.

\section{Performance runs}
\label{sec:performance}
We evaluate the computational performance of the coupled ICESEE-ISSM framework through both strong and weak scaling experiments using the fully parallelized configuration described in \Sec{fully_parallelization}. Our objective is to assess the scalability and efficiency of ICESEE when coupled with a large-scale ice sheet model, where memory requirements and I/O operations are dominant, and where ICESEE cannot easily override MPI communicators from the host model (ISSM) during coupling. All simulations are performed on the \textbf{Phoenix} \citep{pace_phoenix} high-performance computing cluster at the Georgia Institute of Technology. Each compute node on Phoenix is equipped with dual Intel Xeon CPUs (24~cores per node) and 192~GB of RAM, interconnected via Mellanox HDR InfiniBand (100-200~Gb/s). 

To isolate the performance of ICESEE itself, we keep the number of ISSM ranks (\textit{model\_nprocs}) fixed throughout the scaling experiments, since our focus is not on evaluating the standalone performance of ISSM but rather the efficiency of ICESEE when integrated with a computationally intensive model. We use the same physical and numerical setup as described in \Sec{issm}, except that the simulation length is reduced to 40~years, sufficiently large to represent a realistic workload while maintaining manageable runtimes for scaling tests.

In both scaling tests, the problem size and ISSM ranks are held fixed while the number of ICESEE ranks varies from 5 to 320. Each ISSM-MATLAB instance uses four ranks, with an additional rank reserved for ICESEE control, resulting in five total ranks per ensemble member. Thus, the total number of MPI ranks in the coupled system is given by
\begin{equation}
    \text{ICESEE ranks} = N_{p} \times (\text{model\_nprocs} + 1),
\end{equation}
where $N_{p}$ is the number of ranks pinned to each ISSM instance.

\subsection{Strong scaling}
\label{sec:strong}
In the strong scaling experiments, both the ensemble size and the problem size are held fixed while the number of ICESEE ranks is varied. The speedup is computed following Amdahl's law~\citep{amdahl} as
\begin{equation}
    S_p = \frac{T(1)}{T(N_p)},
\end{equation}
where $T(1)$ is the wall-clock time using a single ICESEE rank executing several ISSM instances sequentially, and $T(N_p)$ is the time taken to complete the same workload using $N_p$ ICESEE ranks. The parallel efficiency is defined as
\begin{equation}
    E_p = S_p \times \frac{N_1}{N_p}.
\end{equation}
As the number of resources increases, the total wall-clock time decreases nearly proportionally up to about 80~ranks, beyond which the speedup and efficiency begin to taper off (Table~\ref{tab:strong_table}). This decline is primarily due to the overhead introduced by the growing number of MATLAB instances and the increasing communication cost between MATLAB and Python during coupling. Although the forecast step in ICESEE is fully parallelized, additional time is still spent during the analysis phase and in parallel I/O operations.

\begin{figure}[h]
\centering
  \includegraphics[width=0.65\linewidth]{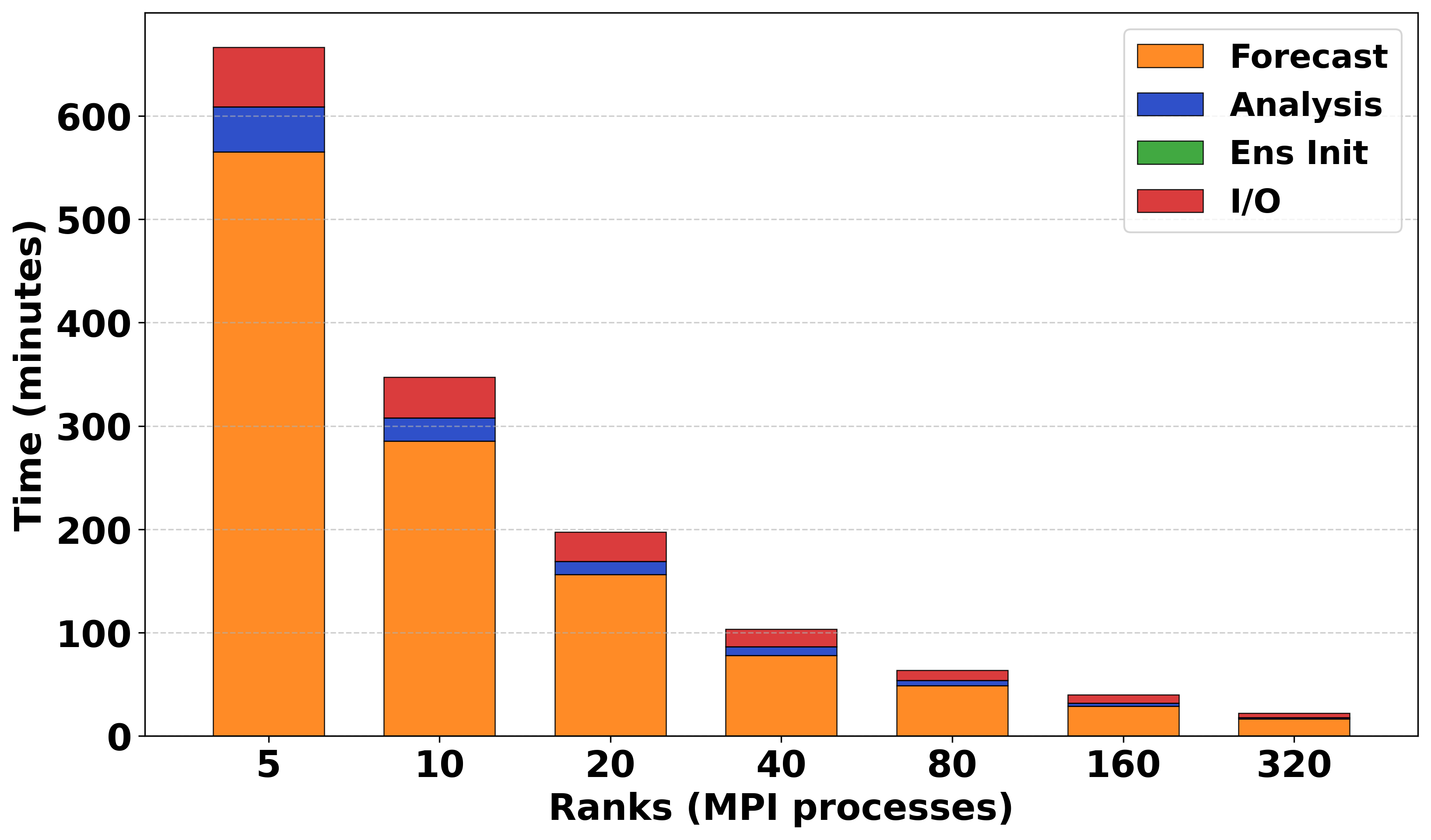}
  \caption{Strong scaling runtime breakdown for the ICESEE-ISSM coupling showing the forecast, analysis, ensemble initialization, and I/O contributions. The forecast step scales most efficiently, while the relative cost of I/O and analysis becomes more noticeable at higher process counts.}
  \label{fig:strong-stacked}
\end{figure}

\begin{table}[ht]
\centering
\caption{Strong scaling performance of ICESEE-ISSM coupling.}
\begin{tabular}{ccccccc}
\toprule
\textbf{Nodes} & \textbf{ICESEE Ranks} & \textbf{ISSM Ranks} & $N_{\text{ens}}$ & \textbf{Wall Time (min)} & \textbf{Efficiency (\%)} & \textbf{Speedup} \\
\midrule
1 & 5 & 4 & 64 & 695.69 & 100.00 & 1.00 \\
1 & 10 & 4 & 64 & 365.01 & 95.30 & 1.91 \\
1 & 20 & 4 & 64 & 206.85 & 84.08 & 3.36 \\
2 & 40 & 4 & 64 & 116.55 & 74.61 & 5.97 \\
4 & 80 & 4 & 64 & 81.31 & 53.47 & 8.56 \\
7 & 160 & 4 & 64 & 44.27 & 49.10 & 15.71 \\
14 & 320 & 4 & 64 & 26.06 & 41.72 & 26.70 \\
\bottomrule
\end{tabular}
\label{tab:strong_table}
\end{table}
\Fig{strong-stacked} provides a detailed breakdown of the total runtime by component. The \textit{forecast} step (orange) dominates computational cost across all configurations, but its time decreases substantially with increasing MPI ranks, confirming effective parallelism. The \textit{analysis} (blue) and \textit{I/O} (red) components also scale reasonably well. However, their relative contribution to the total runtime increases at higher process counts due to synchronization and communication overheads. Generally, the results demonstrate strong scalability and efficient utilization of computational resources across the tested range.

\subsection{Weak scaling}
\label{sec:weak-scaling}
For the weak scaling experiments, we increase both the problem size and the number of ensemble members proportionally so that the workload per ICESEE rank remains constant. Specifically, we begin with 4 ensemble members and 5 ICESEE ranks, keeping the number of ISSM ranks fixed. The number of ICESEE ranks is then doubled each time the number of ensemble members is doubled, following the weak scaling model proposed by \citet{Gustafson}. This configuration allows us to evaluate how the ICESEE-ISSM coupling performs as the total problem size grows while maintaining a fixed computational load per rank. The weak-scaling efficiency is computed as
\begin{equation}
    E_{\text{weak}} = \frac{T(1)}{T(N_p)},
\end{equation}
where $T(1)$ is the wall-clock time for the ICESEE-ISSM baseline configuration: 5~ranks and 4~ensemble members, and $T(N_p)$ is the wall-clock time obtained when scaling up to $N_p$ ICESEE ranks. 

\begin{table}[t]
\centering
\caption{Weak scaling performance of ICESEE--ISSM coupling.}
\begin{tabular}{cccccc}
\toprule
\textbf{Nodes} & \textbf{ICESEE Ranks} & \textbf{ISSM Ranks} & $N_{\text{ens}}$ & \textbf{Wall Time (min)} & \textbf{Efficiency (\%)} \\
\midrule
1 & 5 & 4 & 4 & 78.56 & 100.00 \\
1 & 10 & 4 & 8 & 81.59 & 96.28 \\
1 & 20 & 4 & 16 & 83.38 & 94.22 \\
2 & 40 & 4 & 32 & 89.92 & 87.36 \\
4 & 80 & 4 & 64 & 97.90 & 80.24 \\
7 & 160 & 4 & 128 & 108.90 & 72.13 \\
14 & 320 & 4 & 256 & 127.87 & 61.43 \\
\bottomrule
\end{tabular}
\label{tab:weak_table}
\end{table}

\begin{figure}[ht]
\centering
  \includegraphics[width=0.6\linewidth]{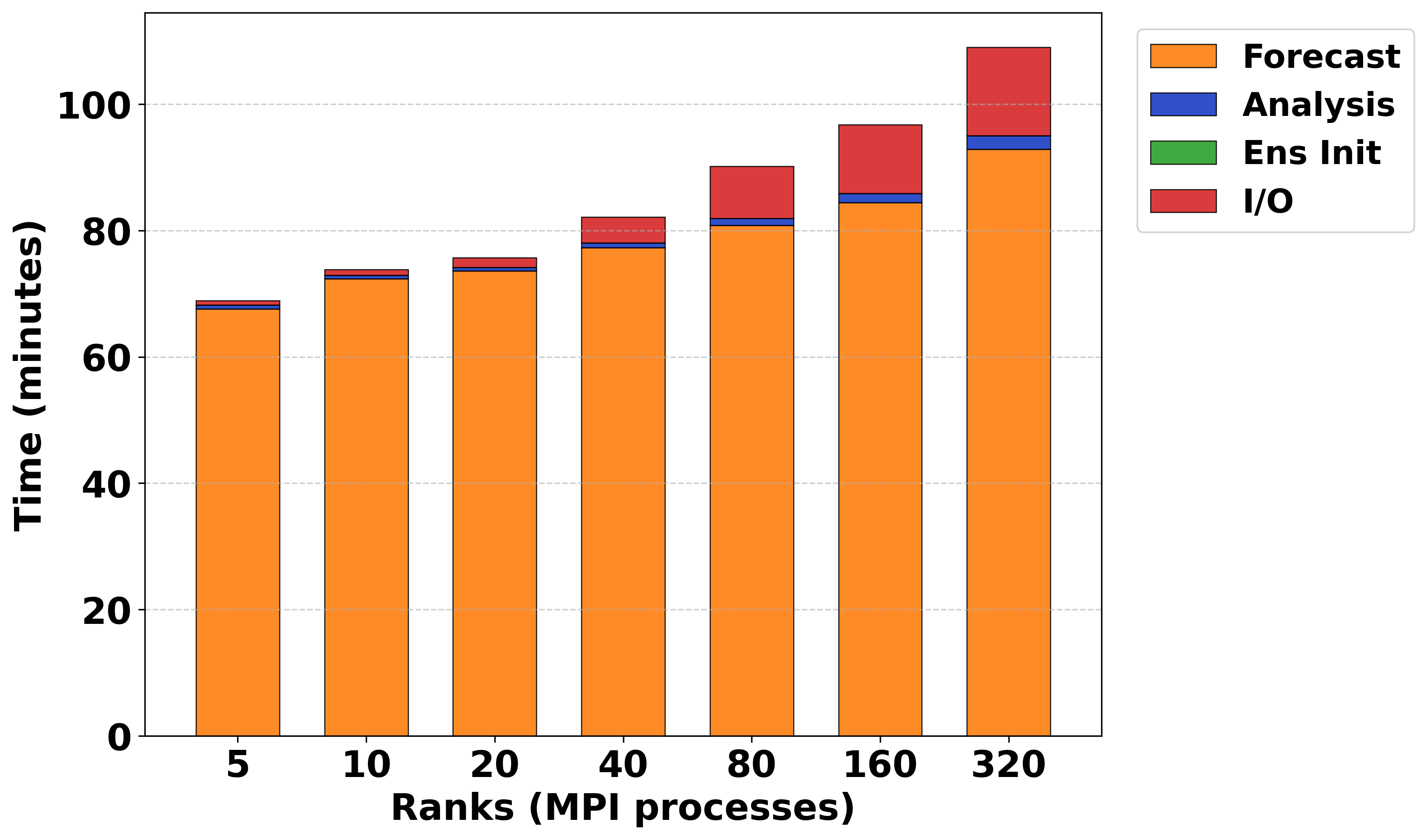}
  \caption{Weak scaling runtime breakdown for the ICESEE--ISSM coupling showing the forecast, analysis, ensemble initialization, and I/O contributions. The forecast phase remains dominant and scales well, while I/O overhead increases slightly at higher process counts due to collective data synchronization.}
  \label{fig:weak-stacked}
\end{figure}
As shown in Table~\ref{tab:weak_table}, the total wall-clock time increases modestly as the number of ranks and ensemble members increases. Starting from 78.56 minutes for five ranks and four ensemble members, the runtime grows to 127.87 minutes at 320 ranks and 256 ensemble members. This gradual increase indicates that ICESEE maintains good parallel scalability for large ensemble sizes. The overall weak-scaling efficiency decreases from 100\% to approximately 61\% as the system scales up, primarily due to communication overheads, I/O synchronization costs among distributed ensemble members, and additional forecast synchronization overheads introduced by the increasing number of concurrent ISSM instances.

Figure~\ref{fig:weak-stacked} shows the breakdown of the total runtime by component. Similar to the strong-scaling results, the \textit{forecast} phase (orange) dominates the total computational cost and scales well across all configurations. The \textit{analysis} (blue) and \textit{ensemble initialization} (green) stages contribute minimally to the overall runtime, while the relative cost of \textit{I/O} (red) increases slightly at higher process counts. The observed growth in I/O time can be attributed to collective data writes and read synchronization across multiple ensemble members and nodes. Despite these communication and I/O effects, ICESEE demonstrates efficient parallel performance under weak scaling, confirming that its design effectively supports large-scale ensemble data assimilation problems.

\section{Conclusion}
\label{sec:conclusion}
ICESEE is a versatile and extensible data assimilation framework designed to enhance the predictive capabilities of ice-sheet models through advanced Ensemble Kalman Filtering techniques and tightly coupled inverse methods. By supporting integration with both Python-based and MATLAB/C++-based models, ICESEE provides a unified platform for real-time state estimation, parameter inference, and forecasting in ice-sheet dynamics. The framework incorporates multiple parallelization strategies, enabling efficient execution on high-performance computing (HPC) systems, and offers flexible coupling mechanisms adaptable to a wide range of modeling architectures.

ICESEE has been successfully applied to two representative ice-sheet models, Icepack and ISSM, demonstrating its adaptability and effectiveness in assimilating synthetic observations to improve model accuracy. The integration with Icepack highlights ICESEE’s ability to interface seamlessly with Python-based modeling workflows, while the coupling with ISSM showcases its capability to operate with MATLAB-based systems through a dedicated MATLAB server. In the ISSM application, ICESEE further enables the indirect estimation of unobserved parameters, such as the basal friction coefficient, by combining assimilated state variables with ISSM’s inverse modeling capabilities. This hybrid assimilation–inversion approach allows ICESEE to exploit cross-variable covariances for state correction while leveraging physics-based cost-function minimization to infer poorly observed parameters, resulting in improved ice geometry, velocity fields, and grounding line evolution.

Performance analyses confirm that ICESEE achieves strong and weak scaling efficiency on HPC systems, underscoring its suitability for large-scale, data-assimilative ice-sheet simulations. Overall, ICESEE represents a significant step toward a unified, high-performance framework that integrates ensemble-based data assimilation with inverse modeling, advancing state estimation, parameter inference, and uncertainty quantification in next-generation glaciological modeling.

\section*{Acknowledgments}
We acknowledge the computing resources that made this work possible provided by the Partnership for an Advanced Computing Environment (PACE) at Georgia Tech in Atlanta, GA, with computing credits provided through a startup from the University System of Georgia. We would like to thank research scientist Fang (Cherry) Liu for her assistance in challenges related to PACE and HPC.

\section*{Financial Support}
The authors and the ICESEE project are supported by Award No. 2235920 from the National Science Foundation Office of Polar Programs.


\vspace{-1.0cm}


\codedataavailability{The current version of ICESEE is available on GitHub at \url{https://github.com/ICESEE-project/ICESEE} under the BSD 2-Clause License. The exact version of the model, including all configuration and plotting scripts used to run and generate the results presented in this paper, is archived on Zenodo at \url{https://doi.org/10.5281/zenodo.18716132} \citep{kyanjo_2026_18716132}.}












\vspace{-1.1cm}

\authorcontribution{
Brian Kyanjo: conceptualization, software development, numerical experiments, and original draft preparation. Talea L. Mayo: guidance on data assimilation methodology and manuscript editing. Alexander Robel: supervision, integration into ice sheet modeling, and manuscript revision. Both Talea L. Mayo and Alexander Robel provided critical feedback and general project oversight.
}

\vspace{-1cm}

\competinginterests{The authors declare no competing interests present} 









\bibliographystyle{copernicus}
\bibliography{literature.bib}

\end{document}